\documentclass[pre,aps,amsmath,amssymb,amsfonts,floatfix,superscriptaddress,showpacs,twocolumn,footinbib]{revtex4-1}
\usepackage{graphicx}
\usepackage{amsmath,amsfonts}
\usepackage{xcolor}
\usepackage{color}
\usepackage{mathtools}
\usepackage{eufrak}
\usepackage{pgfplots}
\usepackage{soul}
\usepackage{multirow}
\usepackage{comment}

\newcommand{\up}{\uparrow}
\newcommand{\dn}{\downarrow}

\newcommand{\ch}{\ensuremath{\text{ch}}}
\newcommand{\sz}{\ensuremath{\text{sp}}}

\newcommand{\pp}{\ensuremath{{pp}}}
\newcommand{\ph}{\ensuremath{{ph}}}
\newcommand{\phv}{\ensuremath{\overline{ph}}}
\newcommand{\trip}{\ensuremath{\text{t}}}
\newcommand{\sing}{\ensuremath{\text{s}}}
\newcommand{\firr}{\ensuremath{\text{firr}}}
\newcommand{\omegap}{\ensuremath{\widetilde{\omega}}}

\usepackage{tikz}

\usetikzlibrary{calc}
\usetikzlibrary{decorations.pathmorphing}
\usetikzlibrary{decorations.pathreplacing}
\usetikzlibrary{decorations.markings}
\usetikzlibrary{shapes.misc}
\usetikzlibrary{shapes}
\usetikzlibrary{positioning}
\usetikzlibrary{snakes}
\usetikzlibrary{arrows}
\usetikzlibrary{fadings}
\usetikzlibrary{calc,arrows}
\tikzstyle{decision} = [diamond, draw, fill=blue!20, text width=4.5em, text badly centered, node distance=3cm, inner sep=0pt]
\tikzstyle{block} = [rectangle, draw, fill=blue!20, text width=7em, text centered, rounded corners, minimum height=5em]
\tikzstyle{line} = [draw, -latex']
\tikzstyle{cloud} = [draw, ellipse,fill=red!20, node distance=3cm, minimum height=2em]
\tikzstyle{overbrace style}=[decorate,decoration={brace,raise=2mm,amplitude=3pt}]
\tikzstyle{overbrace text style}=[font=\footnotesize, above, pos=.5, yshift=3mm]
\tikzset{snake it/.style={decorate, decoration=snake}}
\usetikzlibrary{3d}
\usepackage{mathtools}
    \tikzset{
            partial ellipse/.style args={#1:#2:#3}{
                        insert path={+ (#1:#3) arc (#1:#2:#3)}
                            }
                        }

\usetikzlibrary{calc}
\tikzset{
            inertial frame/.style = {x={(-20:2cm)}, y={(-160:2cm)}, z={(90:2cm)}},
              local frame/.style = {shift={(local origin)}, x={(40:.7cm)}, y={(150:.7cm)}, z={(105:.7cm)}}
          }

    \tikzset{middlearrow/.style={
                decoration={markings,
                            mark= at position 0.65 with {\arrow{#1}} ,
                                    },
                                            postaction={decorate}
                                                }
                                                }
\tikzset{cross/.style={cross out, draw, 
         minimum size=2*(#1-\pgflinewidth), 
                  inner sep=0pt, outer sep=0pt}}

\def\presuper#1#2%
  {\mathop{}%
   \mathopen{\vphantom{#2}}^{#1}%
   \kern-0.5\scriptspace%
   #2}

\usepackage{hyperref}
\hypersetup{colorlinks=true,breaklinks,linkcolor=blue,urlcolor=blue,citecolor=blue}
\begin{document}

    \pgfmathdeclarefunction{gauss}{2}{%
          \pgfmathparse{1/(#2*sqrt(2*pi))*exp(-((x-#1)^2)/(2*#2^2))}%
          }
    \pgfmathdeclarefunction{mgauss}{2}{%
          \pgfmathparse{-1/(#2*sqrt(2*pi))*exp(-((x-#1)^2)/(2*#2^2))}%
          }
    \pgfmathdeclarefunction{lorentzian}{2}{%
        \pgfmathparse{1/(#2*pi)*((#2)^2)/((x-#1)^2+(#2)^2)}%
          }
    \pgfmathdeclarefunction{mlorentzian}{2}{%
        \pgfmathparse{-1/(#2*pi)*((#2)^2)/((x-#1)^2+(#2)^2)}%
          }

\author{Friedrich Krien}
\affiliation{International School for Advanced Studies (SISSA) and CNR-IOM, Via Bonomea 265, I-34136, Trieste, Italy}
\author{Angelo Valli}
\affiliation{Institute for Solid State Physics, Vienna University of Technology, 1040 Vienna, Austria}
\affiliation{International School for Advanced Studies (SISSA) and CNR-IOM, Via Bonomea 265, I-34136, Trieste, Italy}
\author{Massimo Capone}
\affiliation{International School for Advanced Studies (SISSA) and CNR-IOM, Via Bonomea 265, I-34136, Trieste, Italy}


\title{Single-boson exchange decomposition of the vertex function}

\begin{abstract}
    We present a decomposition of the two-particle vertex function of the single-band Anderson impurity model which imparts a physical interpretation of the vertex
    in terms of the exchange of bosons of three flavors. We evaluate the various components of the vertex for an impurity model
    corresponding to the half-filled Hubbard model within dynamical mean-field theory.
    For small values of the interaction almost the entire information encoded in the vertex function corresponds to single-boson exchange processes,
    which can be represented in terms of the Hedin three-leg vertex and the screened interaction. Also for larger interaction,
    the single-boson exchange still captures scatterings between electrons and the dominant low-energy fluctuations
    and provides a unified description of the vertex asymptotics. The proposed decomposition of the vertex does not
    require the matrix inversion of the Bethe-Salpeter equation.
    Therefore, it represents a computationally lighter and hence more practical alternative to the parquet decomposition.
\end{abstract}

\maketitle
\section{Introduction}

Feynman's diagrammatic technique provides us with an intuitive, yet mathematically rigorous representation
of interacting quantum many-body systems in terms of the elementary physical processes which arise at the various orders of perturbation theory.
At the same time, the Feynman diagrams convey an effective information in terms of renormalized {quasi}-particles, dressed collective excitations, and their interaction.
The single-particle self-energy can be interpreted as a frequency- and momentum-dependent self-consistent field~\cite{Bickers04} experienced by the particles as a consequence of their mutual interaction.
Its analytical properties often have consequences for observables, for example,
kinks can mark an intermediate-energy regime of a Fermi liquid~\cite{Byczuk06},
or at a Mott metal-insulator transition the self-energy even diverges~\cite{Georges96}.

On the other hand, two-particle correlations lead to a more complex network of intertwined particle-hole and particle-particle scatterings, which undermines an interpretation in terms of effective quantities with a direct physical content. 
Although we can define a formal analog to the single-particle self-energy, the two-particle self-energy~\footnote{
  {We use the notion `two-particle self-energy'~\cite{Pruschke96} for the Bethe-Salpeter kernel,
  which is the vertex irreducible with respect to pairs of Green's functions in one particular channel
  (hor./vert. particle-hole or particle-particle).}}, it does not have an obvious interpretation in terms of an effective field.
This makes it difficult to use this object for the formulation of approximate schemes driven and supported by intuition.
Nevertheless, the two-particle self-energy plays a key role in the formulation of conserving approximations~\cite{Baym62}
and it is used to compute the full two-particle scattering amplitude (full vertex) through the parquet equations~\cite{Dominicis64,Dominicis64-2}.
The latter can be used directly as the starting point for approximations~\cite{Bickers91,Bickers04,Toschi07},
but they are also useful to justify simplified schemes~\cite{Janis08,Janis17}, such as the ladder approximation~\cite{Moriya85,Toschi07,Rohringer12}.

The traditional formulation of diagrammatic theories based on two-particle quantities poses however significant challenges.
One of them is the immense algorithmic complexity of the parquet equations~\cite{Eckhardt18},
which restricts direct numerical applications to impurity models~\cite{Chen92,Janis08,Rohringer12,Janis17} and small clusters~\cite{Valli15,Schueler17,Kauch17,Pudleiner19,Kauch19,Kauch19-2}.
Furthermore, the perturbation theory based on the two-particle self-energy has revealed fundamental problems which have received much attention in the context of non-perturbative approaches, such as the dynamical mean-field theory (DMFT)~\cite{Georges96}.
Here, already at moderate coupling strength the two-particle self-energy shows divergences~\cite{Schaefer13,Schafer16,Chalupa18}
which signal a breakdown of the perturbation theory and a multi-valued Luttinger-Ward functional~\cite{Gunnarsson17}.

It must be however noticed that, at the present state, 
these divergences appear primarily as a technical/mathematical problem that arises from the
inversion of the generalized susceptibility matrix, whose eigenvalues can cross zero~\cite{Schaefer13,Schafer16,Chalupa18,Gunnarsson17,Vucicevic18,Thunstroem18}.
Even though it is reasonable to expect an influence of the divergences on observables~\cite{Gunnarsson17,Nourafkan19}, no clear and general physical interpretation has been given for this phenomenology.
{From the point of view of the parquet decomposition, the divergences give rise to irregular
and oscillating contributions to the single-particle self-energy via the equation of motion~\cite{Gunnarsson16}.
This poses serious difficulties to the algorithms for solving self-consistently the parquet equations in the non-perturbative regime.} Indeed, the perturbation series is not absolutely convergent~\cite{Kozik15},
which hence leaves open the possibility that the parquet decomposition corresponds to an ill-behaved rearrangement of a conditionally convergent infinite series~\cite{Gunnarsson17}. 
These are indications that the partitioning of the full vertex defined by the parquet equations is not in general useful to define effective quantities. Interestingly, useful information can however be extracted from this representation when it is possible to identify an effective particle~\cite{Kauch19}.
On the other hand, the full vertex function itself \textit{does} retain the physical meaning of an effective interaction~\cite{Krien19-2} which implies that, for example, it allows a fluctuation diagnostic of the single-particle self-energy~\cite{Gunnarsson15,Wu17,Gunnarsson18,Antipov19}.

In view of these difficulties it seems crucial to identify an alternative partitioning of the full vertex which identifies effective quantities, allowing for a simpler physical interpretation.
Ideally, the building blocks of this representation should be the collective excitations mentioned in the first lines of this manuscript. If the correct excitations are identified, one expects that, in a given regime, only some characteristic excitations  prevail while the others are intrinsically weak.
The contribution of these excitations to the vertex diagrams is then either large or small, respectively, without any complicated cancelation between them.
In this alternative scheme the divergences of the two-particle self-energy should be avoided. 

In this work we present a decomposition of the vertex function along these lines, expressing the latter in terms of effective exchange bosons.
The coupling between renormalized fermionic excitations and the bosons is mediated by the Hedin three-leg vertex~\cite{Hedin65}.
This leads to a decomposition of the full vertex into four components, similar to the parquet decomposition,
which allows us to relate each main feature of the vertex to characteristic scattering events that involve the exchange of a \textit{single} boson.
{This identification unifies various parametrizations of the vertex function used in the context of $GW$ approaches~\cite{Giuliani05,Held11}, the functional renormalization group (fRG)~\cite{Karrasch08,Husemann09}, partial bosonizations~\cite{Wetterich07,Wetterich10,Stepanov16-2,Stepanov18},
as well as in the treatment of vertex asymptotics~\cite{Wentzell16,Kaufmann17} in parquet approaches~\cite{Li16,Kauch17}
and in the calculation of the DMFT susceptibility~\cite{Kunes11,Tagliavini18,Krien19}.
However, the novelty of our approach resides in the fact that the \textit{single-boson exchange} (SBE) decomposition
is an exact representation of the full vertex which allows for a simple physical interpretation of its features in terms of boson exchange processes. Moreover, with the introduction of the Hedin vertex, one can formulate a set of \textit{exact} parquet-like equations for the full vertex~\cite{Krien19-3}. In contrast to the standard parquet approach, the SBE decomposition does not require matrix inversions and only relies on algebraic expressions involving physical response functions, which are manifestly non-divergent, unless a symmetry is spontaneously broken. Thus, the absence of matrix inversions and the representation in terms of the Hedin vertex strongly reduce the algorithmic complexity compared to the standard parquet formalism.
The representation of the vertex in terms of physical correlation functions is reminiscent of the dual fermion and dual boson approaches~\cite{Rubtsov08,Rubtsov12}.
}

An ideal test case for the SBE decomposition is the Anderson impurity model (AIM), where an impurity site with a local Hubbard repulsion is hybridized with a non-interacting bath. The zero-dimensional character of the model simplifies the treatment, while the model remains nontrivial. The AIM also allows for numerically exact solutions for the vertex function using
continuous-time quantum Monte Carlo (QMC) solvers~\cite{Gull11,Wallerberger19} with improved estimators~\cite{Hafermann12,Gunacker16,Kaufmann19} which can be used to benchmark our approach.
Furthermore, the AIM can be connected to the Hubbard model on the lattice via the DMFT mapping.
Therefore, in the following Sec.~\ref{sec:aimvertex} we define the DMFT approximation and the corresponding 
vertex function of the auxiliary AIM. In Sec.~\ref{sec:mainresult} we present in words our main result.
The rigorous derivation of the SBE decomposition follows in Sec.~\ref{sec:sbe}.
In Sec.~\ref{sec:results} we present exemplary numerical results, we conclude in Sec.~\ref{sec:conclusions}.

\section{Vertex function of the auxiliary Anderson impurity model}\label{sec:aimvertex}
We build the AIM which DMFT associates to the half-filled Hubbard model on the square lattice in the paramagnetic state
\begin{align}
    H = &-\sum_{\langle ij\rangle\sigma}{t}_{ij} c^\dagger_{i\sigma}c^{}_{j\sigma}+ U\sum_{i} n_{i\up} n_{i\dn},\label{eq:hubbard}
\end{align}
where ${t}_{ij}$ is the nearest neighbor hopping between lattice sites $i,j$, its absolute value ${t}=1$ is the unit of energy.
$c^{},c^\dagger$ are the annihilation and creation operators, $\sigma=\up,\dn$ the spin index.
$U$ is the Hubbard repulsion between the densities $n_{\sigma}=c^\dagger_{\sigma}c^{}_{\sigma}$.
The action of the auxiliary Anderson impurity model of DMFT is defined as,
\begin{align}
  S_{\text{AIM}}=&-\sum_{\nu\sigma}c^*_{\nu\sigma}(\imath\nu+\mu-\Delta_\nu)c^{}_{\nu\sigma}+U \sum_\omega n_{\up\omega} n_{\dn\omega},
  \label{eq:aim}
\end{align}
where $c^*,c$ are Grassmann numbers, $\nu$ and $\omega$ are fermionic and bosonic Matsubara frequencies, respectively,
and $\Delta_\nu$ is the self-consistent hybridization function of DMFT. The chemical potential  is fixed to $\mu=\frac{U}{2}$ in order to enforce particle-hole symmetry which implies  half-filling.
Summations over Matsubara frequencies $\nu, \omega$ contain implicitly the factor $T$, the temperature.
In DMFT the hybridization function is fixed via the self-consistency condition, $G_{ii}(\nu)=g(\nu)$,
where $G_{ii}$ is the local lattice Green's function of the Hubbard model~\eqref{eq:hubbard} in DMFT approximation,
and $g_\sigma(\nu)=-\langle c^{}_{\nu\sigma}c^*_{\nu\sigma}\rangle$ is the Green's function of the AIM~\eqref{eq:aim}.
Since we consider the paramagnetic case the spin label $\sigma$ is suppressed where unambiguous.

\begin{figure*}
\begin{center}
  \includegraphics[width=\textwidth]{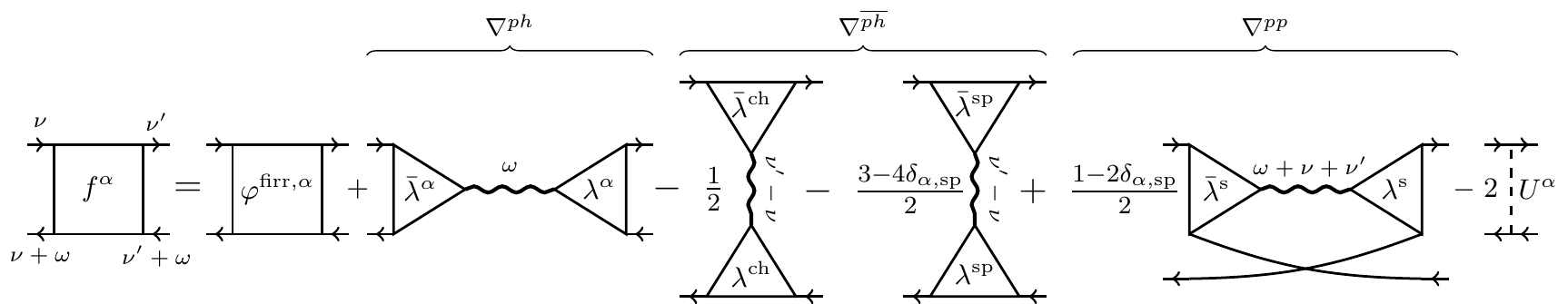}
\end{center}
    \caption{\label{fig:jib} The single-boson exchange decomposition of the vertex function $f^{\ch,\sz}$.
    Triangles denote the Hedin vertices $\lambda^{\alpha}$, wiggly lines the screened interaction $w^\alpha$.
    Diagrams on the right-hand-side from left to right:
    (i) The fully $U$-irreducible four-point vertex $\varphi^{\firr}$.
    (ii) $U$-$ph$-reducible diagrams.
    (iii+iv) $U$-$\overline{ph}$-reducible diagrams.
    (v) $U$-$pp$-reducible diagrams.
    (vi) Double-counting correction that cancels two times the bare interaction $U^\alpha$ (dashed line).
    }
    \end{figure*}

Central to this work is the vertex function, which is the connected part of the four-point correlation function,
\begin{align}
    g^{(4),\alpha}_{\nu\nu'\omega}=&-\frac{1}{2}\sum_{\sigma_i}s^\alpha_{\sigma_1'\sigma_1^{}}s^\alpha_{\sigma_2'\sigma_2^{}}
    \langle{c^{}_{\nu\sigma_1}c^{*}_{\nu+\omega,\sigma_1'}c^{}_{\nu'+\omega,\sigma_2}c^{*}_{\nu'\sigma_2'}}\rangle\notag,
\end{align}
where $s^\alpha$ are the Pauli matrices and the label $\alpha=\ch,\sz$ denotes the charge and spin channel, respectively.
One obtains the four-point vertex function $f$ by subtracting the disconnected parts from $g^{(4)}$ and amputating four Green's function legs,
\begin{align}
f^\alpha_{\nu\nu'\omega}=&\frac{g^{(4),\alpha}_{\nu\nu'\omega}-\beta g_\nu g_{\nu+\omega}\delta_{\nu\nu'}
    +2\beta g_\nu g_{\nu'}\delta_{\omega}\delta_{\alpha,\ch}}{g_\nu g_{\nu+\omega}g_{\nu'}g_{\nu'+\omega}}\label{eq:4pvertex}.
\end{align}
The vertex function is depicted diagrammatically on the left-hand-side of Fig.~\ref{fig:jib}.
It is a two-particle scattering amplitude, which describes the propagation and interaction of two fully dressed fermionic particles~\cite{Rohringer12}.
In this work we choose the particle-hole notation, where the two left (right) entry points of the vertex correspond to
a particle with energy $\nu$ ($\nu'$) and a hole with energy $\nu+\omega$ ($\nu'+\omega$).

\section{Main result}\label{sec:mainresult}
In this section we introduce our main result, which will be derived in the next Section \ref{sec:sbe}.
\subsection{SBE decomposition}
The main result consists of the {\it exact} diagrammatic decomposition of the vertex function $f$ of the AIM~\eqref{eq:aim}
depicted diagrammatically in Fig.~\ref{fig:jib}, which defines the various contributions to the expansion for the two channels $\alpha=\ch, \sz$
\begin{align}
    f^\alpha=\varphi^{\firr,\alpha}+\nabla^{ph,\alpha}+\nabla^{\overline{ph},\alpha}+\nabla^{\pp,\alpha}-2U^\alpha\label{eq:jib},
\end{align}
where the wiggly lines denote the {\it{dressed}} (or screened) interaction and $\bar{\lambda}^\alpha$ and $\lambda^\alpha$ are three-leg Hedin vertices.

As we shall detail in Sec. \ref{sec:sbe}, the rigorous principle by which the vertex is decomposed is the notion of \textit{reducibility with respect to the bare interaction $U^\alpha$}. A diagram is $U^\alpha$-reducible if it can be split into two parts by removing a bare interaction line $U^\alpha$, 
where $U^\alpha$ is the bare interaction for the charge, spin, and singlet particle-particle channel, defined respectively as,
\begin{align}
    U^\ch=+U,\;\;\;U^\sz=-U,\;\;\;U^\sing=+2U.\label{eq:bareint}
\end{align}
In the triplet particle-particle channel the bare interaction vanishes (see Ref.~\cite{Bickers89} and Sec.~\ref{sec:upp}).

Equation~\eqref{eq:jib} has also a rather transparent physical picture which we describe in the following. 
The contributions on its right-hand-side are, from left to right: A fully $U$-{irreducible} vertex $\varphi^{\firr}$,
a horizontally $U$-{reducible} vertex $\nabla^{ph}$, a vertically $U$-reducible $\nabla^{\overline{ph}}$,
and $\nabla^{\pp}$ which is $U$-reducible in a particle-particle sense.
Lastly, a double-counting correction subtracts twice the bare interaction $U^\alpha$ that is included in each of the three vertices $\nabla$.

The decomposition~\eqref{eq:jib} allows an interpretation of the vertex function $f$ in terms of processes involving the exchange of effective bosons.
The boson lines appearing in the different contributions correspond to the three generic fluctuations of the single-band AIM which are connected to the bosonic operators
$\rho^\ch=n_\up+n_\dn, \rho^\sz=n_\up-n_\dn$, and $\rho^-=c_\up c_\dn$, that is, charge (ch), spin (sp), and singlet particle-particle (s) fluctuations, respectively.
In Fig.~\ref{fig:jib} each vertex $\nabla$ connects two of its corners with fermionic entry/exit points to the other two corners via a bosonic line, the screened interaction,
\begin{eqnarray}
    w^\alpha(\omega)=U^\alpha+\frac{1}{2}U^\alpha\chi^\alpha(\omega)U^\alpha.\label{eq:w}
\end{eqnarray}
Here, $\chi^\alpha$ is the susceptibility, which is the correlation function of the bosonic operator with the corresponding 
flavor $\alpha=\ch,\sz,\sing$. The bosonic fluctuations lead to a screening of the bare interaction~\eqref{eq:bareint}.

The screened interaction is enclosed by right- and left-sided Hedin vertices $\bar{\lambda}^\alpha$ and $\lambda^\alpha$.
These three-legged vertices are formally similar to the fermion-boson vertex,
which conveys information about the response of the fermionic spectrum to an \textit{external} applied field~\cite{vanLoon18,Krien19-2}.
{On the other hand, in Hedin's reformulation of the equation of motion~\cite{Hedin65} the Hedin vertex plays the role of a coupling of the fermions
to the \textit{internal} bosonic fluctuations of the system}, which are also of electronic origin~\footnote{
    The Hedin (or `proper') vertex differs from the fermion-boson vertex precisely by the dielectric function~\cite{Giuliani05}.}.
As shown in the results [see Fig.~\ref{fig:hedin}], the Hedin vertex is in general of order unity, $|\lambda|\approx1$.
For the qualitative discussion of the SBE decomposition~\eqref{eq:jib} we can therefore ignore the energy-dependence of the fermion-boson coupling as a first approximation.

\begin{figure}[b]
    \begin{center}
  \includegraphics[width=.48\textwidth]{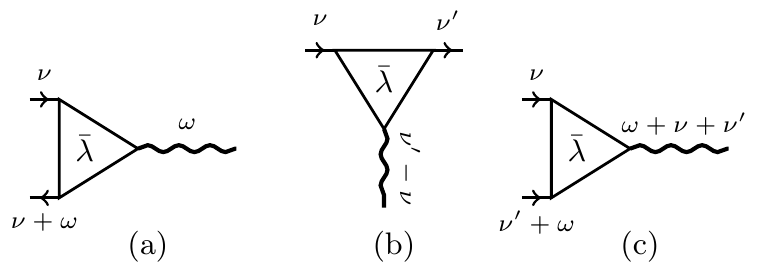}
\end{center}
    \caption{\label{fig:events} Boson exchanges of the SBE decomposition in Fig.~\ref{fig:jib}.
    (a) Particle-hole annihilation.
    (b) Particle propagation and boson exchange with a hole.
    (c) Particle-particle annihilation.
    }
    \end{figure}

We can interpret the horizontally $U$-reducible vertex $\nabla^{ph}$ as follows.
The right-sided Hedin vertex $\bar{\lambda}$ represents the \textit{annihilation} of a particle-hole pair with energies $\nu$ and $\nu+\omega$ [see also Fig.~\ref{fig:events} (a)].
In turn, the left-sided $\lambda$ represents the \textit{creation} of a particle-hole pair with energies $\nu'$ and $\nu'+\omega$,
and in these events a boson with energy $\omega$ is exchanged.
As a consequence, fermionic particles do in fact \textit{not} propagate from left to right,
and the vertex $\nabla^{ph}$ therefore merely captures annihilation and creation of particle-hole pairs.
This process is resonant for an energy $\omega\approx0$ of the boson -- which is independent of the fermionic energies $\nu$ and $\nu'$.
Since the Hedin vertices are in general of order unity,
the fermionic frequencies may be varied without changing drastically the magnitude of $\nabla^{ph}$.
Therefore, this vertex contributes a constant background to the full vertex function $f$~\cite{Krien19}.
 
Next, we consider the vertically $U$-reducible diagrams represented by $\nabla^{\overline{ph}}$.
This vertex, despite being similar to $\nabla^{ph}$, has a different physical interpretation~\footnote{
    {The interpretation of $\nabla^{ph}$ and $\nabla^{\overline{ph}}$ depends on the \textit{`frame of reference'},
    which is fixed by the (horizontal) particle-hole notation in Eq.~\eqref{eq:4pvertex} with the transferred frequency $\omega$.
    One may also adopt a vertical notation with the transferred frequency $\nu'-\nu$.
    Then $\nabla^{ph}$ and $\nabla^{\overline{ph}}$ interchange their roles of particle-hole annihilation/creation and propagation, respectively, since they are related by the crossing-symmetry (cf. Sec.~\ref{sec:uvph}).}
}.
This is clear from Fig.~\ref{fig:events} (b), which shows that the Hedin vertex components of $\nabla^{\overline{ph}}$
represent the \textit{propagation} of particle and hole, respectively, and their exchange of a boson.
This process is resonant for a small energy transfer $\nu'-\nu\approx0$ and hence the vertex $\nabla^{\overline{ph}}$ depends sensitively on the fermionic frequencies.
    
We turn to the contribution of the particle-particle channel, $\nabla^{{pp}}$.
In this case the Hedin vertex can be interpreted in terms of the annihilation (creation) of a particle-particle pair, as depicted in Fig.~\ref{fig:events} (c).
The resonance of this contribution lies at $\nu+\nu'+\omega\approx0$, which is the sum of the particle energies.

The last term of our decomposition is the fully $U$-irreducible vertex $\varphi^{\firr}$. This part of the full vertex function can not be represented in terms of single-boson exchange,
therefore, it describes \textit{multi}-boson exchange processes and events that can not be represented in terms of boson exchange at all.
In Sec.~\ref{sec:results} we will however show that $\varphi^{\firr}$ decays for large $\nu$ or $\nu'$ in all directions, which is in fact a general property of this vertex and we will discuss the approximation where this term is neglected.
This will show that the single-boson exchange processes described by the vertices $\nabla$ capture substantial information about the full vertex function $f$.
Henceforth, we shall refer to equation~\eqref{eq:jib} as a \textit{single-boson exchange (SBE) decomposition} of the full vertex and to $\nabla$ as SBE vertices.

\subsection{Comparison with other approaches}

We summarize some crucial features of the SBE decomposition and compare it to established methods to treat electronic correlations at the two-particle level.

Beside the SBE formalism, in diagrammatic theories it is natural to represent 
contributions to vertex functions in terms of fluctuations (bosons) of electronic origin.
For example, they are the eponymous feature of the fluctuation exchange approximation (FLEX)~\cite{Bickers89}.
In particular, diagrams with the structure of the SBE vertices $\nabla$ with \textit{one} wiggly line are commonly known as Maki-Thompson diagrams~\cite{Maki68,Thompson70}, 
whereas the Aslamasov-Larkin diagrams~\cite{Aslamasov68}, which correspond to \textit{two-boson} exchange processes,
are included in $\varphi^{\textit{firr}}$, see also Refs.~\cite{Glatz11,Tsuchiizu16,Marenko04}.
The novel aspect of the SBE decomposition is however a reclassification of \textit{all} diagrams for the full vertex
according to the picture of a single exchanged boson, leading to three reducible classes and one fully irreducible class, similar to the parquet decomposition.
Moreover, given the exact (or an approximate) fully irreducible vertex $\varphi^{\textit{firr}}$
the SBE vertices $\nabla$ can be reconstructed from a set of self-consistent equations, analogous to the parquet equations.
The resulting self-consistent {SBE equations for the Hedin vertex} will be discussed elsewhere.

At first sight, the SBE decomposition may appear simply as an alternative, with clear formal similarities, to the parquet decomposition.
There are however at least three important advantages of the SBE decomposition:

\textit{First}, as discussed above, the SBE vertices correspond to diagrams that give rise to the high frequency structures of the full vertex, whereas $\varphi^{\textit{firr}}$ contains diagrams which decay fast in frequency; the SBE vertices $\nabla$ therefore naturally recover the vertex asymptotics~\cite{Wentzell16}.
This property is not shared by the parquet decomposition, where the various contributions have no characteristic high-frequency behavior~\cite{Rohringer12}.

\textit{Second}, the SBE vertices are given by the Hedin vertex $\lambda$ and the screened interaction $w$,
which can be expressed in terms of physical correlation functions.
Therefore, these quantities can be computed directly through their Lehmann representations~\cite{Tagliavini18} or using QMC techniques. 
On the other hand, the two-particle self-energy $\Gamma$ can not be computed in a similar way.
In fact, $\Gamma$ is obtained via inversion of the associated Bethe-Salpeter equation,
which may be non-invertible, giving rise to divergences of $\Gamma$~\cite{Schaefer13,Schafer16,Chalupa18,Gunnarsson17,Vucicevic18,Thunstroem18}. 
This problem is absent in the SBE, where the inversion of the Bethe-Salpeter equation is neither required to obtain the SBE vertices $\nabla$ in an exact calculation,
nor in a self-consistent reconstruction of $\nabla$ from $\varphi^{\textit{firr}}$. 
The SBE decomposition is therefore completely unaffected by singular behavior of the two-particle self-energy. 

\textit{Third}, due to the structure of the SBE vertices $\nabla$, their computational cost is drastically reduced with respect to the reducible vertices of the parquet decomposition. 
Furthermore, it also suggests numerically inexpensive approximations to the full vertex,
such as neglecting the fully irreducible vertex $\varphi^{\text{\firr}}$, 
and neglecting the frequency dependence of the Hedin vertex (i.e., $\lambda^{\alpha}\approx\pm1$).
These two schemes realize two well-defined physical approximations, in which the the four-point vertex is given in terms of three-point 
or even only two-point vertices, respectively.
{In fact, the latter option recovers a weak-coupling parametrization of the vertex function~\cite{Karrasch08,Gunnarsson15},
which the SBE decomposition unifies with asymptotic expressions for the vertex~\cite{Wentzell16,Tagliavini18} into an exact framework.}
A similar strategy is not accessible within the standard parquet decomposition 
{since the contributions to the asymptotic belong to objects 
with different reducible properties.} 

\renewcommand{\arraystretch}{2}
\begin{table}[t]
\begin{tabular}{ c l c }
  \hline
  \hline
  \textcolor{white}{acronym} & approximation & \multicolumn{1}{c}{SBE formalism} \\ 
  \hline
2OPT & $f=U$         & $\varphi^{\firr}=0$;  
                       $\nabla^{ph},\nabla^{\overline{ph}},\nabla^{pp} = U$\\
GW   & $\Sigma=gw$   & \multirow{2}{*}{ \hspace{-0.2in}\Bigg{\}} 
                       $\varphi^{\firr}=0$;  
                       $\nabla^{\overline{ph}}, \nabla^{pp}  = U$;  
                       $\nabla^{ph} = w$ }\\
RPA  & $\Gamma=U$  \\  
  \hline
    $\textit{PA}^{*}$         & $\Lambda=U$             & $\varphi^{\firr}=0$ \\
D$\Gamma$A$^{*}$ & $\Lambda=\Lambda_{imp}$ & $\varphi^{\firr}=\varphi^{\firr}_{imp}$ \\
  \hline
  \hline
\end{tabular}
\caption{\label{tab:approximations} Different approximations redefined within the SBE formalism.
    The second-order perturbation theory (2OPT), the GW and the RPA can be recovered within the SBE formalism. 
    $^{*}$For parquet-based schemes, such as the parquet approximation (PA) and the D$\Gamma$A,
    one can define \textit{analogous} approximations within the SBE formalism,
    which have however a different diagrammatic content (see text).}
\end{table}
\begin{figure}[b]
    \begin{center}
      \includegraphics[width=.48\textwidth]{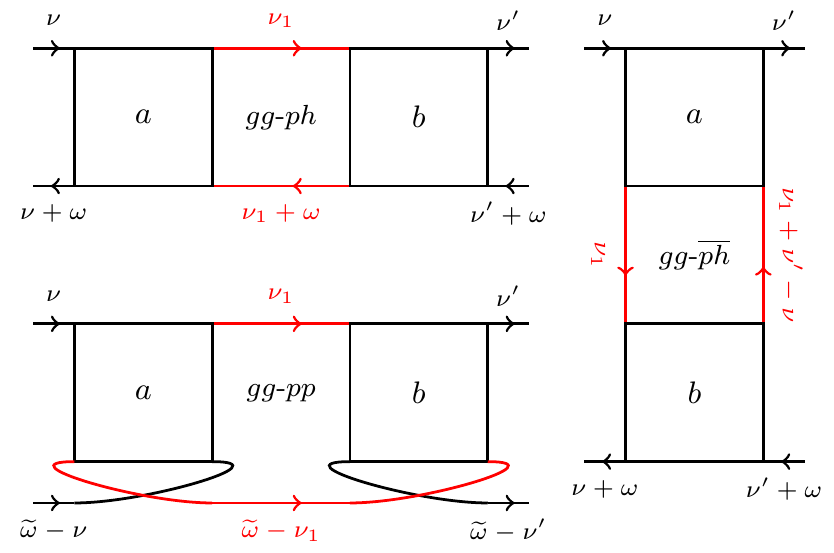}
    \end{center}
    \vspace{-.3cm}
    \caption{\label{fig:ggred}
    $gg$-reducible diagrams: The blocks $a$ and $b$ can be separated by cutting the red lines.
    The transferred frequencies of the particle-hole channels are $\omega$ (top) and $\nu'-\nu$ (right),
    for the particle-particle channel it is $\omegap=\omega+\nu+\nu'$ (bottom). Green's function lines do \textit{not} belong to $a$ or $b$.
    }
\end{figure}
\begin{figure}[b]
    \begin{center}
      \includegraphics[width=.48\textwidth]{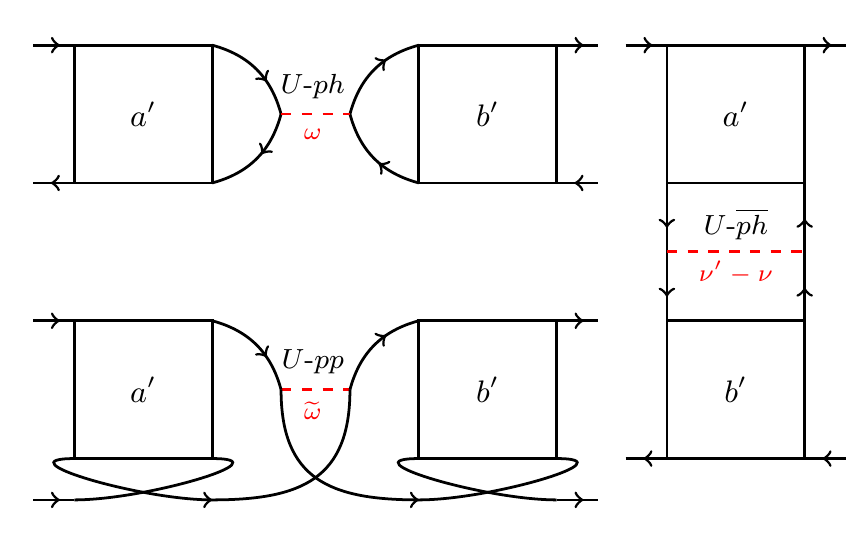}
\end{center}
    \vspace{-.2cm}
    \caption{\label{fig:ured}
    $U$-reducible diagrams: The blocks $a'$ and $b'$ are separated by removing a bare interaction (red dashed line).
    Green's function lines \textit{do} belong to $a'$ or $b'$, which can be equal to $1$. The case $a'=b'=1$ leads to the bare interaction.\\
    }
    \end{figure}

In this respect it is interesting to make a connection between the SBE formalism 
and established diagrammatic theories, which is summarized in Tab.~\ref{tab:approximations}. 
For instance, the second-order perturbation theory (2OPT) is obtained at the two-particle level 
by approximating the full vertex $f$ with the bare interaction. 
In the SBE formalism this corresponds to neglecting the fully irreducible vertex, $\varphi^{\firr}=0$, 
and the frequency dependence of the SBE vertices $\nabla^{ph,\alpha},\nabla^{\overline{ph},\alpha},\nabla^{pp,\alpha}= U^{\alpha}$.
Also the vertex corrections to the self-energy of the $GW$ approach are naturally recovered in the present framework
by setting $\varphi^{\textit{\firr}}=0$, $\nabla^{\overline{ph},\alpha}$, $\nabla^{pp,\alpha}=U^\alpha$ and $\nabla^{ph,\alpha}=w^\alpha$.~\footnote{
    Setting $\nabla^{\overline{ph}}$ and $\nabla^{pp}$ to the bare interaction cancels the double counting correction in Eq.~\eqref{eq:jib},
  $\nabla^{ph}=w$ implies that the Hedin vertices of the particle-hole channel are set to unity, $\lambda^\ch=\lambda^{\sz}=1$.}
Hence, the full vertex is given by the screened interaction, $f^\alpha=w^\alpha$.
This highlights how within $GW$ the {vertex corrections} in the particle-hole channel 
are absorbed into the screened interaction via the Hedin equation~\cite{Hedin65}, however,
the contribution of the vertices in the other channels are not treated on equal footing. 
The same approximation in the SBE formalism also leads to the random-phase approximation (RPA) in the $\ph$ channel for the susceptibility.
Analogous resummations in the other channels can be recovered retaining either $\nabla^{\overline{\ph}}$ or $\nabla^{\pp}$.

Within the standard parquet formalism, one can build theories by taking different approximations
to the fully irreducible vertex, often called $\Lambda$~\cite{Rohringer12}. 
The simplest example is the parquet \textit{approximation}~\cite{Bickers91,Bickers04,Toschi07},
where $\Lambda$ is approximated by the bare interaction, $\Lambda=U$. 
A more sophisticated approximation for a lattice model is the dynamical vertex approximation (D$\Gamma$A),
where the fully irreducible vertex of the lattice is approximated with the local one of the impurity,
$\Lambda_{\textit{lat}}=\Lambda_{\textit{imp}}$~\cite{Toschi07,Valli15}.
We can define {analogous} approximations within the SBE formalism, 
where the fully irreducible vertex is either set to $\varphi^{\firr}=0$ (SBE \textit{approximation}),~\footnote{
    In the SBE decomposition the bare interaction is included in the SBE vertices $\nabla$,
    the simplest but nontrivial approximation to the fully irreducible vertex is therefore $\varphi^{\firr}=0$.
}
or to the one of the impurity $\varphi^{\firr}=\varphi^{\firr}_{imp}$ (SBE-D$\Gamma$A). 
However, as the diagrammatic content of $\varphi^{\textit{\firr}}$ is \textit{not} the same as the one of $\Lambda$,
these two approximation schemes are not equivalent to their parquet counterparts.

\section{Single-boson exchange decomposition}\label{sec:sbe}
In this section we derive the SBE decomposition~\eqref{eq:jib}.

\subsection{Two notions of reducibility}\label{app:red}
The SBE decomposition is based on the notion of $U$-reducibility, see, for example, Ref.~\cite{Giuliani05}.
This concept can be defined starting from the more common definition of reducibility with respect to pairs of Green's functions. Every diagram for the full vertex $f$ can either be cut into two parts by removing two Green's function lines ($gg$-reducible) or it is fully ($gg$-)irreducible.
Furthermore, each $gg$-reducible diagram can be cut in only one of three ways, either by removing a particle-hole pair horizontally ($gg$-$ph$) or vertically ($gg$-$\overline{ph}$),
or by removing a particle-particle pair ($gg$-$pp$). This is shown in Fig.~\ref{fig:ggred}, see also Refs.~\cite{Rohringer12,Bickers04}.

\begin{figure}[t]
    \begin{center}
      \includegraphics[width=.48\textwidth]{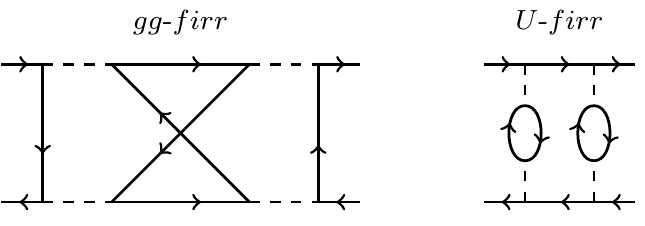}
    \end{center}
    \caption{\label{fig:firr}
    (Left) A fully $gg$-irreducible 'envelope' diagram. This diagram is also fully $U$-irreducible.
    (Right) A fully $U$-\textit{irreducible} but $gg$-${ph}$-\textit{reducible} diagram (belongs to top white area in Fig.~\ref{fig:venn}).
    Note that without either of the bubble insertions the diagram would be $U$-$ph$-reducible.
    }
\end{figure}

\begin{figure}
    \begin{center}
      \includegraphics[width=.43\textwidth]{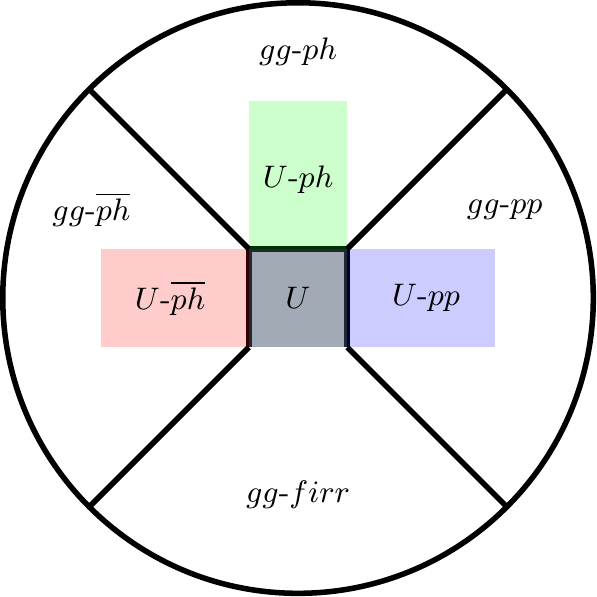}
\end{center}
    \caption{\label{fig:venn} Venn diagram for the $gg$- and $U$-reducible classes of diagrams for the vertex function.
    Areas enclosed by black lines symbolize the $gg$-reducible (top, left, right) and the fully $gg$-irreducible (bottom) classes
    (i.e., the traditional parquet decomposition).
    The square region in the middle belongs to $gg$-$firr$ and contains only the bare interaction.
    The classes of $U$-reducible diagrams are marked in colors,
    they lie within the respective $gg$-reducible class, except for the bare interaction,
    which is element of $U$-$ph$, $U$-$\overline{ph}$, and $U$-$pp$.
    The four white areas comprise the fully $U$-irreducible diagrams.
    }
\end{figure}

In Fig.~\ref{fig:ured} we show $U$-reducible diagrams. The latter can be cut into two parts by removing a bare interaction line $U^\alpha$.
There are again two particle-hole channels ($U$-$ph$ and $U$-$\overline{ph}$) and one particle-particle channel ($U$-$pp$).
Due to the Green's function lines connected to the bare interaction, each $U$-reducible diagram is $gg$-reducible in one and the same channel.
Therefore, a diagram is either $U$-reducible in only one of the three channels or it is fully $U$-irreducible.
Notice that the converse is in general false, as $gg$-reducible diagrams can be fully $U$-irreducible, e.g., the right diagram in Fig.~\ref{fig:firr}. 
Furthermore, the fully $gg$-irreducible ($gg$-$firr$) and the fully $U$-irreducible ($U$-$firr$) classes are unrelated,
that is, fully $U$-irreducible diagrams can be $gg$-reducible or fully $gg$-irreducible.
Fully irreducible diagrams are shown in Fig.~\ref{fig:firr}. 
There is however one exception, the bare interaction itself, which is $U$-reducible in all three channels, but fully $gg$-irreducible. 
The two notions of reducibility are represented as a Venn diagram in Fig.~\ref{fig:venn}. 

It follows that with proper care of the double-counting of the bare interaction, we can write the full vertex $f$
as the sum of the $U$-reducible ($\nabla$) and fully $U$-irreducible ($\varphi^\firr$) diagrams, which reads in a full frequency notation,
\begin{align}
    f^\alpha_{\nu\nu'\omega}\!=\!\varphi^{\firr,\alpha}_{\nu\nu'\omega}\!+\!\nabla^{ph,\alpha}_{\nu\nu'\omega}
    \!+\!\nabla^{\overline{ph},\alpha}_{\nu\nu'\omega}\!+\!\nabla^{\pp,\alpha}_{\nu\nu',\omega+\nu+\nu'}\!-\!2U^\alpha.\label{eq:jib_full}
\end{align}
We subtracted the bare interaction two times, which is counted once by each vertex $\nabla$.
The $U$-$pp$-reducible vertex $\nabla^{\pp}$ is denoted in the particle-particle notation with transferred frequency $\tilde{\omega}=\omega+\nu+\nu'$.
In the following we derive the SBE vertices $\nabla$ one by one.

\subsection{$\mathbf{U}$-$\mathbf{ph}$ channel}\label{sec:uph}
The horizontally $U$-reducible diagrams $\nabla^{ph}$ are those commonly associated to the Hedin formalism~\cite{Hedin65},
and they can be represented in terms of the Hedin vertex, as shown in Refs.~\cite{Giuliani05,Held11} and more recently in Ref.~\cite{Krien19}.
The Hedin vertex is related to a response function, the three-point (fermion-boson) correlation function~\cite{vanLoon18,Krien19-2},
\begin{align}
    g^{(3),\alpha}_{\nu\omega}=&\frac{1}{2}\sum_{\sigma\sigma'}s^\alpha_{\sigma'\sigma}
    \langle{c^{}_{\nu\sigma}c^{*}_{\nu+\omega,\sigma'}\rho^\alpha_\omega}\rangle\notag,
\end{align}
where $\rho^\ch=n_\up+n_\dn$ and $\rho^\sz=n_\up-n_\dn$ are the charge and spin densities and $s^\alpha$ is a Pauli matrix.
The right-sided Hedin vertex (the bosonic end-point is on the right) is defined as ($\langle n\rangle=\langle\rho^\ch\rangle$),
\begin{align}
    \bar{\lambda}^{\alpha}_{\nu\omega}=\frac{g^{(3),\alpha}_{\nu\omega}+\beta g_\nu \langle n\rangle\delta_{\omega}\delta_{\alpha,\ch}}
    {g_\nu g_{\nu+\omega}(1+\frac{1}{2}U^\alpha\chi^\alpha_\omega)},\label{eq:hedinvertex}
\end{align}
where $\chi^\alpha_\omega=-\langle{\rho^\alpha_{-\omega}\rho^\alpha_\omega}\rangle+\beta\langle n\rangle\langle n\rangle\delta_\omega\delta_{\alpha,\ch}$
is the susceptibility in the channel $\alpha$.
One further defines a left-sided Hedin vertex $\lambda$ (with bosonic end-point on the left~\cite{Krien19}).
Under time-reversal symmetry and SU($2$) symmetry the left- and right-sided Hedin vertices are equal,
$\bar{\lambda}=\lambda$, see also Appendix~\ref{app:sym:ph}.

The crucial aspect of the Hedin vertex $\lambda$ is that the factor $1+\frac{1}{2}U^\alpha\chi^\alpha_\omega$ in the denominator of Eq.~\eqref{eq:hedinvertex}
removes the horizontally $U$-reducible diagrams that belong to the class $U$-$ph$~\cite{Hertz73}, $\lambda$ is therefore $U$-$ph$-\textit{irreducible}~\cite{Rohringer16}.
We can use $\lambda$ to separate all $U$-$ph$-reducible diagrams from the full vertex function,
\begin{align}
    f^\alpha_{\nu\nu'\omega}=&\varphi^{\ph,\alpha}_{\nu\nu'\omega}+\nabla^{ph,\alpha}_{\nu\nu'\omega}\label{eq:fph}.
\end{align}
Here, $\varphi^{\ph}$ denotes a $U$-$ph$-\textit{irreducible} vertex and $\nabla^{ph}$ is the $U$-$ph$-\textit{reducible} vertex in Eq.~\eqref{eq:jib_full}.
As shown in Ref.~\cite{Krien19}, $\nabla^{ph}$ is given by the Hedin vertex,
\begin{align}
    \nabla^{ph,\alpha}_{\nu\nu'\omega}=\bar{\lambda}^{\alpha}_{\nu\omega}\,w^\alpha_\omega\,\lambda^{\alpha}_{\nu'\omega}.\label{eq:nablahph}
\end{align}
This is shown as the second diagram on the right-hand-side of Fig.~\ref{fig:jib}.
The Hedin vertices are connected at their bosonic end-points via the screened interaction defined in Eq.~\eqref{eq:w}.
Notably, equation~\eqref{eq:fph} plays an analogous role for the SBE decomposition as the Bethe-Salpeter equation
of the horizontal particle-hole channel does for the parquet decomposition.

\subsection{$\mathbf{U}$-$\mathbf{\overline{ph}}$ channel}\label{sec:uvph}
The vertical particle-hole channel is conceptually not different from the horizontal one and
$\nabla^{\overline{ph}}$ in Eq.~\eqref{eq:jib_full} can be obtained using
the crossing symmetry of the full vertex function in the paramagnetic system~\cite{Rohringer12},
\begin{align}
    f^{\alpha}_{\nu\nu'\omega}
    \!=&-\!\frac{1}{2}\!\left(f^{\ch}_{\nu,\nu+\omega,\nu'-\nu}\!+\![3\!-\!4\delta_{\alpha,\sz}]f^{\sz}_{\nu,\nu+\omega,\nu'-\nu}\right),\label{eq:crossing}
\end{align}
where $\alpha=\ch,\sz$.
The crossing symmetry tells how the vertex $f$ acts when the particle-hole scatterings are not
considered to happen horizontally with transferred frequency $\omega$, but vertically with transferred frequency $\nu'-\nu$.
As Eq.~\eqref{eq:crossing} shows, the flavor labels $\ch, \sz$ are not conserved by the crossing relation.
Viewed vertically the charge vertex $f^\ch$ has a spin component and $f^\sz$ has a charge component.

When we apply the crossing relation~\eqref{eq:crossing} to Eq.~\eqref{eq:fph} we obtain the following \textit{new} relation,
\begin{align}
    f^\alpha_{\nu\nu'\omega}=&\varphi^{\phv,\alpha}_{\nu\nu'\omega}+\nabla^{\overline{ph},\alpha}_{\nu\nu'\omega}\label{eq:fvph},
\end{align}
where $\varphi^{\phv}$ is a $U$-$\overline{ph}$-\textit{irreducible} vertex and the $U$-$\overline{ph}$-\textit{reducible} diagrams are given as ($\alpha=\ch,\sz$),
\begin{align}
    \nabla^{\overline{ph},\alpha}_{\nu\nu'\omega}
    \!=&-\!\frac{1}{2}\!\!\left(\!\nabla^{ph,\ch}_{\nu,\nu+\omega,\nu'-\nu}\!+\![3\!-\!4\delta_{\alpha,\sz}]\nabla^{ph,\sz}_{\nu,\nu+\omega,\nu'-\nu}\!\right)\!.\!\label{eq:nablavph}
\end{align}
The $U$-$\overline{ph}$-\textit{reducible} diagrams are therefore simply obtained from the crossing relation~\eqref{eq:crossing},
they are shown as the third and fourth diagram on the right-hand-side of Fig.~\ref{fig:jib}.
Nevertheless, $\nabla^{\overline{ph}}$ and $\nabla^{{ph}}$ generate different diagrams, which we show explicitly in Appendix~\ref{app:u2}.

\subsection{$\mathbf{U}$-$\mathbf{{pp}}$ channel}\label{sec:upp}
We now come to the particle-particle channel, which requires some further preparation.
First, we need to define a suitable three-point correlation function,
\begin{align}
    g^{(3),\sing}_{\nu\widetilde{\omega}}=\left\langle c_{\nu\up}c_{\widetilde{\omega}-\nu,\dn}\rho^{+}_{\widetilde{\omega}}\right\rangle.\label{eq:g3pp}
\end{align}
Here, we introduced the pair density $\rho^+=c^\dagger_\up c^\dagger_\dn$ and its conjugate $\rho^-=c_\dn c_\up$.
The index $\sing$ indicates the \textit{singlet} pairing channel and $\widetilde{\omega}$ is the transferred frequency of a particle-particle pair.

We further define a right-sided Hedin-like vertex,
\begin{align}
    \bar{\lambda}^{\sing}_{\nu\omegap}=\frac{g^{(3),\sing}_{\nu\omegap}}{g_\nu g_{\omegap-\nu}(1+\frac{1}{2}U^s\chi^\sing_{\omegap})},\label{eq:lambdasing}
\end{align}
where the denominator includes the respective bare interaction $U^s$ and the singlet pairing susceptibility
$\chi^\sing_{\omegap}=-\left\langle \rho^-_{-\omegap}\rho^{+}_{\omegap}\right\rangle$.
As for the particle-hole case, one defines a left-sided vertex $\lambda^\sing$,
which is equal to $\bar{\lambda}^\sing$ under time-reversal and SU($2$) symmetry, see Appendix~\ref{app:sym:pp}.
For simplicity, we will also refer to $\bar{\lambda}^{\sing}$ as a Hedin vertex, although it is not part of the original formalism~\cite{Hedin65}.

Note that in the $pp$-channel the bare interaction $U^s$ differs from the $ph$-channel by a factor $2$ due to the indistinguishability of identical particles~\cite{Bickers89,Rohringer12},
\begin{align}
    U^s=2U.\label{eq:ubarepp}
\end{align}
One further defines a singlet vertex function $f^s$, it is obtained from the particle-hole vertices as~\cite{Rohringer12},
\begin{align}
    f^{\sing}_{\nu\nu'\omegap}=&\frac{1}{2}\left(f^{\ch}_{\nu\nu',\omegap-\nu-\nu'}-3f^{\sz}_{\nu\nu',\omegap-\nu-\nu'}\right),\label{eq:fsinglet}
\end{align}
where $\omegap=\omega+\nu+\nu'$ is the transferred frequency of particle-particle pairs.
As we did above for the $ph$- and $\overline{ph}$-channels, the $U$-$pp$-reducible diagrams will now be separated from the vertex $f^{\sing}$
and expressed in terms of $\lambda^{\sing}$. We show in Appendix~\ref{app:upp:firr} that in full analogy to the particle-hole channels we can write,
\begin{align}
    f^\sing_{\nu\nu'\omegap}=&\varphi^{\pp,\sing}_{\nu\nu'\omegap}+\bar{\lambda}^{\sing}_{\nu\omegap}w^\sing_{\omegap}{\lambda}^{\sing}_{\nu'\omegap},
    \label{eq:us}
\end{align}
where $\varphi^{\pp,\sing}$ is $U$-$pp$-irreducible and the screened interaction $w^s$ for the singlet channel appears.
It is given simply by extending the definition~\eqref{eq:w} to the case $\alpha=\sing$.

Eq.~\eqref{eq:us} achieves the separation of $U$-$pp$-reducible diagrams from $f^{\sing}$, however,
the desired $U$-$pp$-reducible contribution is $\nabla^{pp}$ in Eq.~\eqref{eq:jib_full} to the vertices $f^\ch$ and $f^\sz$ in particle-hole notation.
Therefore, we like to write analogous to Eqs.~\eqref{eq:fph} and~\eqref{eq:fvph},
\begin{align}
    f^{\alpha}_{\nu\nu'\omega}=&\varphi^{\pp,\alpha}_{\nu\nu',\omega+\nu+\nu'}+\nabla^{\pp,\alpha}_{\nu\nu',\omega+\nu+\nu'},\label{eq:uspp}
\end{align}
for $\alpha=\ch,\sz$. $\varphi^{\pp,\ch/\sz}$ is a $U$-$pp$-irreducible vertex for the respective channel and we show in Appendix~\ref{app:upp:nablapp} that the $U$-$pp$-reducible diagrams are given as,
\begin{align}
    \nabla^{\pp,\alpha}_{\nu\nu'\omegap}=\frac{1-2\delta_{\alpha,\sz}}{2}\bar{\lambda}^{\sing}_{\nu\omegap}w^\sing_{\omegap}\lambda^{\sing}_{\nu'\omegap},
    \label{eq:nablapp}
\end{align}
which is the fifth diagram on the right-hand-side of Fig.~\ref{fig:jib}.

Lastly, we explain why it is not necessary to consider the triplet-$pp$-channel. One defines the triplet vertex as,
\begin{align}
    f^{\trip}_{\nu\nu'\omegap}=&\frac{1}{2}\left(f^{\ch}_{\nu\nu',\omegap-\nu-\nu'}+f^{\sz}_{\nu\nu',\omegap-\nu-\nu'}\right).\label{eq:triplet}
\end{align}
This vertex represents the interaction of two particles with the same spin ($f^\trip=f^{\up\up}$, see Ref.~\cite{Rohringer12}).
However, this implies that there are no $U$-$pp$-reducible diagrams in $f^{\trip}$,
because the bare Hubbard interaction $Un_{\up}n_{\dn}$ only acts between particles with opposite spin flavors,
therefore, $f^{\trip}$ is by construction $U$-$pp$-\textit{irreducible}~\footnote{
    In multi-orbital Hubbard models off-diagonal matrix-elements of the bare interaction may contribute to the triplet channel,
    which requires to introduce a triplet Hedin vertex $\lambda^\text{t}$ and a screened interaction $w^\text{t}$.
    See also Appendix E of Ref.~\cite{Krien19}.}.

This completes the SBE decomposition.
One should note that Eqs.~\eqref{eq:jib_full},~\eqref{eq:fph},~\eqref{eq:fvph}, and~\eqref{eq:uspp} form a set of parquet-like equations,
which can be closed provided an input for the fully $U$-irreducible vertex $\varphi^{\firr}$~\footnote{
    At the single-particle level the Green's function can be renormalized via the Hedin equation for the self-energy $\Sigma$~\cite{Hedin65,Giuliani05,Held11}.
}.

\section{Numerical examples}\label{sec:results}
In this section we present numerical results for the AIM~\eqref{eq:aim} to verify the exactness of  the SBE decomposition~\eqref{eq:jib_full} and to analyze the weight of the different terms. 
Using continuous-time QMC solvers, we compute the full vertex $f$ and the Hedin vertices $\lambda$ of the impurity model corresponding to self-consistent DMFT calculations for the \textit{half-filled} Hubbard model~\eqref{eq:hubbard} on the square lattice. We then evaluate the SBE vertices $\nabla$ and the fully irreducible vertex $\varphi^{\firr}$ numerically exactly.
The inverse temperature is fixed to $\beta t=5$.
    
The impurity model was solved using two different QMC solvers.
The impurity Green's function $g$, the vertex function $f$, the Hedin vertices $\lambda^\ch$ and $\lambda^\sz$,
and the screened interactions $w^\ch$ and $w^\sz$ of the particle-hole channels were evaluated
using the ALPS solver~\cite{ALPS2} with improved estimators~\cite{Hafermann12}.
The Hedin vertex $\lambda^\sing$ and the screened interaction $w^\sing$ of the particle-particle channel
were obtained using the worm sampling of the w2dynamics package~\cite{Wallerberger19}. We verified that the two solvers yield consistent results.
\begin{figure}
    \begin{center}
      \includegraphics[width=.48\textwidth]{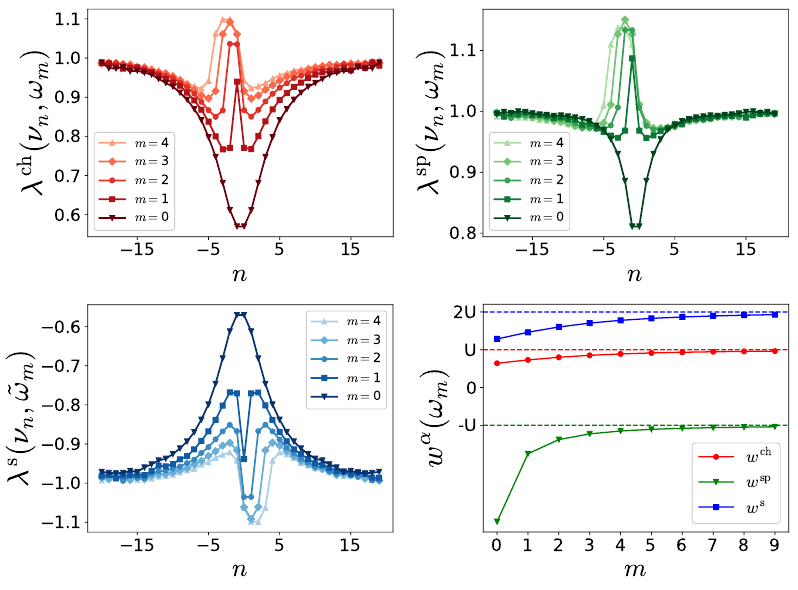}
\end{center}
    \vspace{-0.3cm}
    \caption{\label{fig:hedin} (Color online) Hedin vertex $\lambda^\alpha$ and screened interaction $w^\alpha$ of the channels $\alpha=\ch, \sz,$ and $\sing$.
    The impurity model corresponds to a DMFT calculation at half-filling for $U/t=4$, where static spin fluctuations dominate.
    }
    \end{figure}
\begin{figure*}
    \begin{center}
      \includegraphics[width=.98\textwidth]{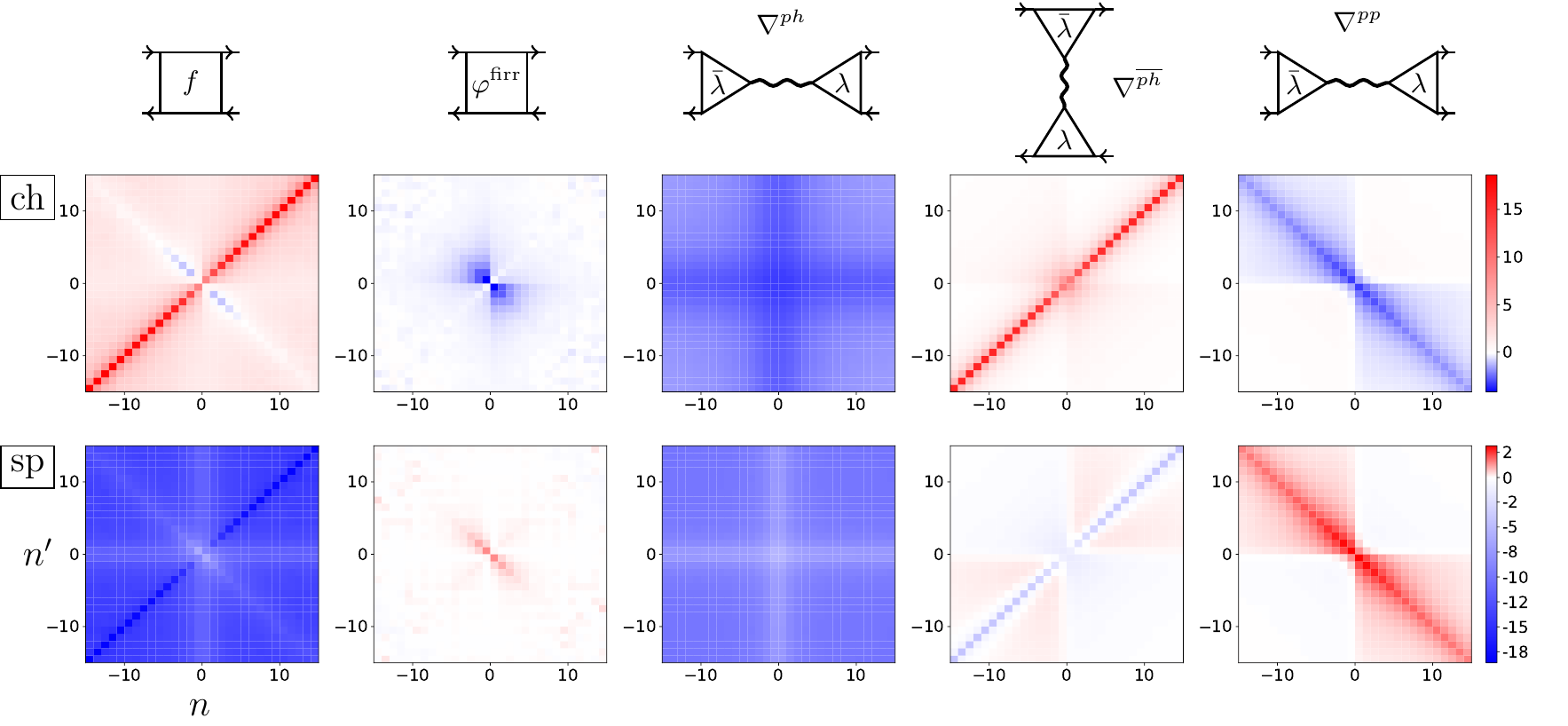}
\end{center}
    \vspace{-0.3cm}
    \caption{\label{fig:examples} (Color online)
    SBE decomposition, see Eq.~\eqref{eq:jib_full} and Fig.~\ref{fig:jib}, for $\alpha=\ch$ (top) and $\alpha=\sz$ (bottom).
    (Left column) Full vertex $f^{\alpha}(\nu_{n\protect\phantom{'}}\!,\nu'_{n'},\omega=0)$ as a function of the Matsubara indices $n, n'$.
    (Third, fourth, and fifth column) SBE vertices $\nabla^{ph}, \nabla^{\overline{ph}}, \nabla^{pp}$. The bare interaction is subtracted from each vertex $\nabla$.
    (Second column) Fully $U$-irreducible vertex $\varphi^{\firr,\alpha}$. 
    All panels of a row share the same color scheme on the right.
    The impurity model corresponds to a DMFT calculation at $U/t=4$.
    }
    \end{figure*}

\subsection{Hedin vertices and screened interaction}\label{sec:results:hedin}
We examine the constituent pieces of the SBE vertices $\nabla$,
the Hedin vertex $\lambda^\alpha$ and the screened interaction $w^\alpha$,
for a DMFT calculation at $U/t=4$ in Fig.~\ref{fig:hedin}.
In the particle-hole channels $\alpha=\ch,\sz$ the Hedin vertex has some structure near $\nu\approx-\omega/2$ and approaches $+1$ for large $|\nu|$ (top panels),
whereas in the singlet particle-particle channel $\alpha=\sing$ the structure lies near $\nu\approx\omegap/2$ and the asymptote is $-1$ (bottom left panel).
In the paramagnetic and particle-hole symmetric AIM the charge and particle-particle fluctuations are related by a symmetry~\cite{Rohringerthesis}.
Correspondingly, the left panels of Fig.~\ref{fig:hedin} show that $\lambda^{\ch}$ and $\lambda^\sing$ are not independent~\footnote{
    In the particle-hole symmetric case we find numerically that $\lambda^\sing_{\nu\omega}=-\lambda^\ch_{\nu,-\omega}$ and $2\chi^\sing(\omega)=\chi^\ch(\omega)$.
    These relations are useful because in some QMC schemes the calculation of the particle-particle quantities is challenging.
    However, without particle-hole symmetry these relations do not hold.}.

At the chosen parameters the three Hedin vertices do not differ much in magnitude, however, the screened interaction $w^\alpha$
indicates that already at $U/t=4$ the spin fluctuations $\alpha=\sz$ are large (see Fig.~\ref{fig:hedin}, bottom right panel).
Near $\omega\approx0$ the screened interaction $w^\sz$ is strongly enhanced beyond its asymptotic value $-U$, hence, bosons with spin flavor dominate.
Due to weak charge and particle-particle fluctuations also $w^\ch$ and $w^\sing$ deviate somewhat from $U$ and $2U$, respectively.

\subsection{SBE vertices}\label{sec:results:components}
The components of the SBE decomposition~\eqref{eq:jib_full} of the static vertex $f^\alpha(\nu,\nu',\omega=0)$
are shown in Fig.~\ref{fig:examples} for $U/t=4$.
The two panels on the left show the full vertex $f$ in the charge and spin channel, the remaining columns show the respective components $\varphi^{\firr}$ and $\nabla$.
To highlight their asymptotic behavior we subtracted the bare interaction $U^\alpha$ from each vertex $\nabla^\alpha$.

The features of the vertex components are consistent with the discussion in Sec.~\ref{sec:mainresult}.
The center panels of Fig.~\ref{fig:examples} show that $\nabla^{ph}$ is indeed largely independent of the fermionic frequencies,
except for $\nu\approx0$ and/or $\nu'\approx0$, where the Hedin vertices depend on the fermionic frequency.
Notice that $\nabla^{ph}$ forms a constant background \textit{beyond} the bare interaction.
The panels in the fourth column of Fig.~\ref{fig:examples} show $\nabla^{\overline{ph}}$,
which has a feature near the main diagonal, $\nu\approx\nu'$, and decays quickly away from it.
On the other hand, $\nabla^{{pp}}$ shown in the right panels of Fig.~\ref{fig:examples} only has a feature near the secondary diagonal, $\nu\approx-\nu'-\omega$.
As expected, both $\nabla^{\overline{ph}}$ and $\nabla^{{pp}}$ do \textit{not} form a constant background beyond the bare interaction.
Finally, the vertex $\varphi^{\firr}$ decays in all directions, it does not have a constant background whatsoever.

Next, we consider the role of the SBE diagrams $\nabla$ for the full vertex $f$.
To this end, we combine them with the double-counting correction in the quantity,
\begin{eqnarray}
    \nabla^{\text{SBE},\alpha}=\nabla^{ph,\alpha}+\nabla^{\overline{ph},\alpha}+\nabla^{\pp,\alpha}-2U^\alpha,
\end{eqnarray}
so that the full vertex is given as $f^\alpha=\varphi^{\firr,\alpha}+\nabla^{\text{SBE},\alpha}$, cf. Eq.~\eqref{eq:jib_full}.
It is interesting to investigate when $\varphi^{\firr,\alpha}$ is small and the corresponding physics.

\begin{figure*}
    \begin{center}
      \includegraphics[width=.98\textwidth]{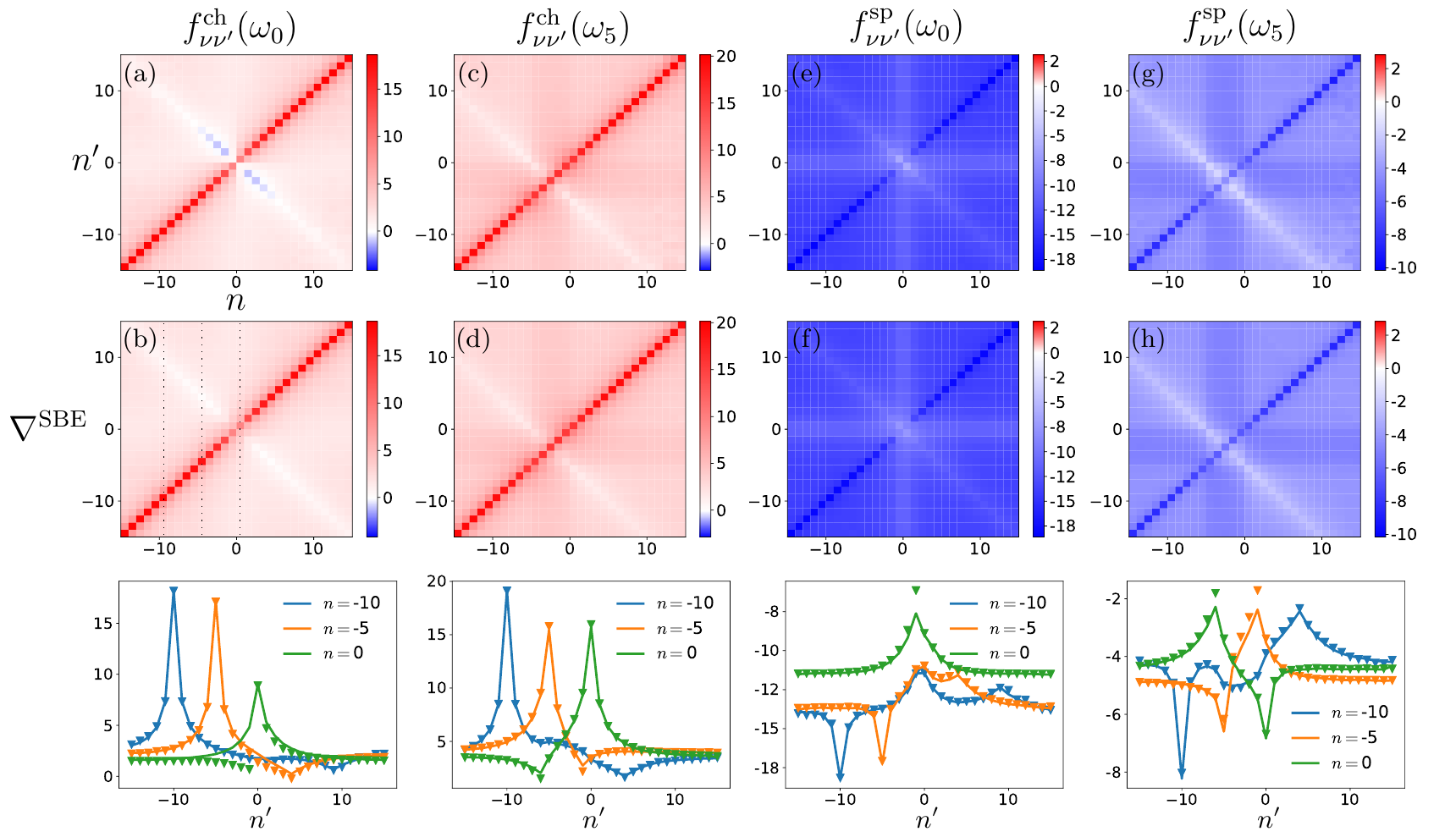}
    \end{center}
    \vspace{-0.3cm}
    \caption{\label{fig:U4} (Color online) Full vertex function $f^\alpha(\nu_{n\protect\phantom{'}}\!,\nu'_{n'},\omega)$ (top row)
    and $\nabla^{\text{SBE},\alpha}(\nu_{n\protect\phantom{'}}\!,\nu'_{n'},\omega)$ (center row)
    as a function of $n, n'$ for fixed $\omega=\omega_0,\omega_5$ and $\alpha=\ch, \sz$.
    (Bottom) Cuts for fixed $\nu=\nu_0, \nu_{-5}, \nu_{-10}$, see also dotted lines in panel (b).
    Triangles show the full vertex $f$, lines show $\nabla^{\text{SBE}}$.
    The impurity model corresponds to a DMFT calculation at $U/t=4$.
    }
\end{figure*}

Once again for $U/t=4$ we show in Fig.~\ref{fig:U4} the full vertex $f$ (top row) and $\nabla^{\text{SBE}}$ (center row).
We focus first on the panels (e)-(h) on the right side of Fig.~\ref{fig:U4} for the spin channel,
which show $f^\sz(\nu,\nu',\omega)$ and $\nabla^{\text{SBE},\sz}(\nu,\nu',\omega)$ for fixed $\omega=\omega_0,\omega_5$ as a function of $\nu$ and $\nu'$.
In fact, there is a perfect qualitative agreement between the vertices both at small and large frequencies [compare panels (e) and (f), as well as (g) and (h), respectively].
Indeed, even quantitative agreement of $f^\sz$ and $\nabla^{\text{SBE},\sz}$ is confirmed by the bottom right panels of Fig.~\ref{fig:U4}, which show slices of the vertices for fixed $\nu=\nu_0,\nu_{-5}$, and $\nu_{-10}$ as function of $\nu'$.
The deviations are small even near $\nu'\approx0$ and $\nabla^{\text{SBE},\sz}$ captures all features of the full vertex $f^\sz$.
Notice that the blue color of the vertices throughout the panels (e)-(h) indicates attractive (negative) interaction in the spin channel, similar to the attractive bare interaction $U^\sz=-U$.

The panels (a)-(d) on the left side of Fig.~\ref{fig:U4} show the charge channel, where the red color indicates repulsive (positive) interaction.
Also here $f^\ch$ and $\nabla^{\text{SBE},\ch}$ mostly agree qualitatively, however, a small region of $f^\ch(\nu,\nu',\omega_0)$ in panel (a) shows \textit{attractive} interaction (blue) for small $\nu\approx-\nu'-\omega$ along the secondary diagonal.
Apparently, the attractive feature of $f^\ch$ is not in $\nabla^{\text{SBE},\ch}$ in panel (b) and indeed it
is instead captured by the fully $U$-irreducible vertex $\varphi^{\firr,\ch}$, which is shown in the second column, top panel of Fig.~\ref{fig:examples}.
This discrepancy for $\omega=\omega_0$ can also be observed in the bottom left panel of Fig.~\ref{fig:U4}.
The comparison to $\omega=\omega_5$ in the neighboring panel shows that the difference $\varphi^{\firr}$
between $f$ and $\nabla^{\text{SBE}}$ quickly decays with the bosonic frequency $\omega$.
The locus of the attractive feature of $f^\ch$ near the secondary diagonal
suggests that it is related to particle-particle scatterings beyond single-boson exchange.

Overall the panels (a)-(h) of Fig.~\ref{fig:U4} show asymptotically perfect agreement of $f$ and $\nabla^{\text{SBE}}$ for large frequencies and mostly good agreement for small frequencies. One should note that the agreement is best, and hence the fully irreducible vertex $\varphi^{\firr,\alpha}$ is small, when the corresponding bosonic fluctuations $\chi^\alpha$ are large. For $U/t=4$ this is the case for the spin channel, as discussed in Sec.~\ref{sec:results:hedin}.
Indeed, Fig.~\ref{fig:examples} shows that $\varphi^{\firr,\sz}$ is very small and confined to a tiny region of $\nu$ and $\nu'$.

\begin{figure*}
    \begin{center}
      \includegraphics[width=.98\textwidth]{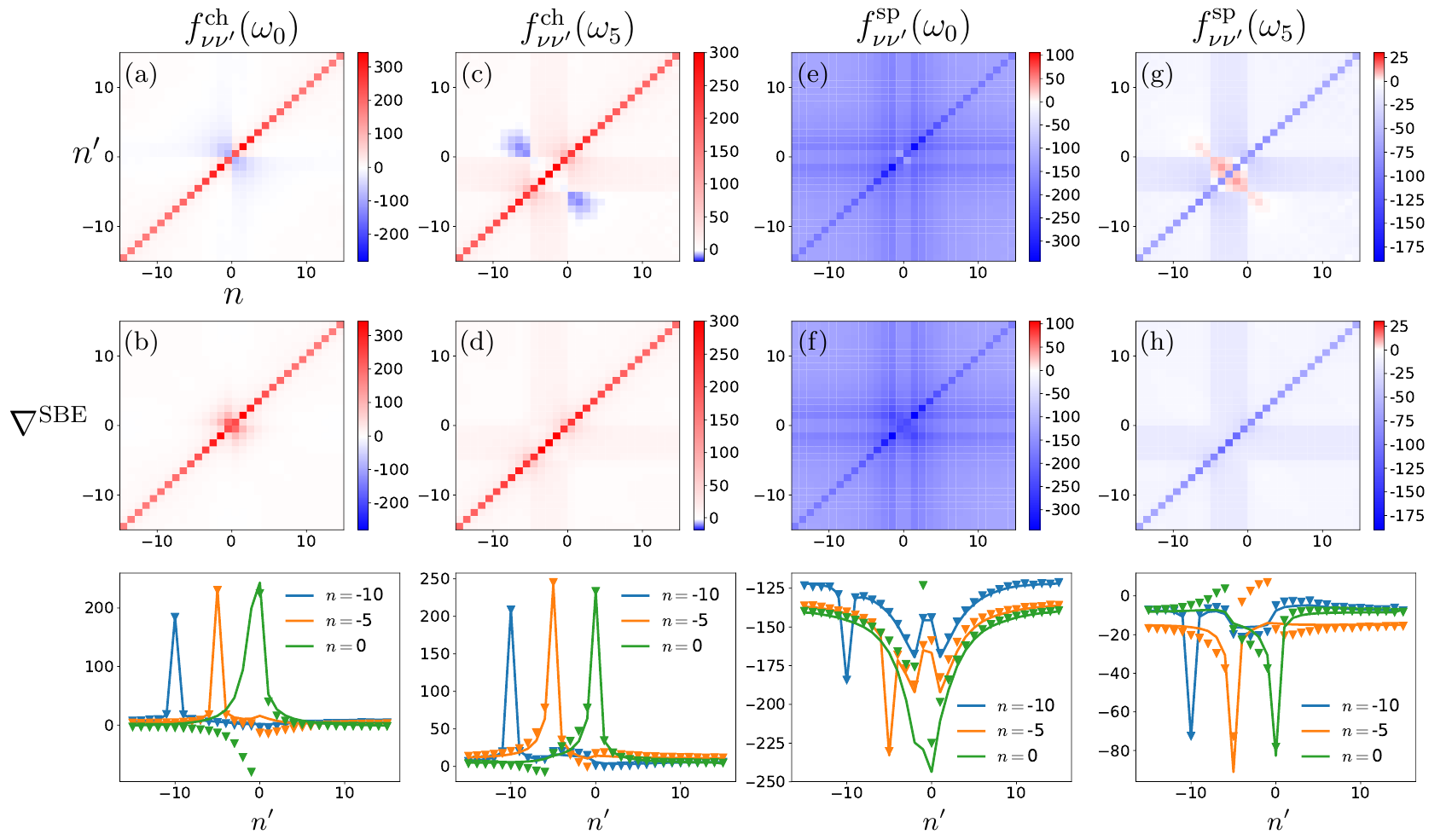}
    \end{center}
    \vspace{-0.3cm}
    \caption{\label{fig:U8} (Color online) The same vertices as in Fig.~\ref{fig:U4} for a DMFT calculation at $U/t=8$.
    }
\end{figure*}

Next, we analyze the SBE vertices for a DMFT calculation at larger Hubbard interaction $U/t=8$, a strongly correlated (bad metal) regime.
The vertices for this case are shown in Fig.~\ref{fig:U8}.
Again, the panels (e) and (f) show good qualitative agreement of $f^\sz$ and $\nabla^{\text{SBE},\sz}$ for $\omega_0=0$,
although the slices in the panel below show some quantitative deviation along the secondary diagonal $\nu\approx-\nu'$.
For finite bosonic frequency $\omega_5$ the spin vertex $f^\sz$ has a repulsive (red) feature on the secondary diagonal $\nu\approx-\nu'-\omega_5$ that is not present in $\nabla^{\text{SBE},\sz}$.

In the charge channel the vertex function $f^\ch$ has two dominant features, see Fig.~\ref{fig:U8} panels (a) and (c),
the main diagonal and an attractive feature on the secondary diagonal $\nu\approx-\nu'-\omega$, similar to $U/t=4$. 
On the other hand, the constant background is small and the secondary diagonal decays for large $\nu, \nu'$.
These features are negligible because the SBE vertices $\nabla^{ph,\ch}$, $\nabla^{\pp,\ch}$, and $\nabla^{\pp,\sz}$
are small in this regime, where the charge and singlet fluctuations $\chi^\ch$ and $\chi^\sing$ are strongly suppressed.

In contrast, the SBE vertex $\nabla^{\overline{ph},\ch}$ of the \textit{vertical} particle-hole channel
is large and contributes the main diagonal in Fig.~\ref{fig:U8} (a)-(d),
because via Eq.~\eqref{eq:nablavph} it has a contribution of the large spin vertex $\nabla^{ph,\sz}$ of the \textit{horizontal} particle-hole channel.
This feedback of the spin fluctuations on charge-charge scatterings is indeed crucial.
As explained in Sec.~\ref{sec:mainresult}, $\nabla^{\overline{ph},\ch}$ describes the \textit{propagation} of particle-hole pairs.
In the bad metal this vertex is strongly repulsive and hence undermines the charge fluctuations.
The large contribution of $\nabla^{ph,\sz}$ to this vertex implies
that charge propagation is suppressed by exchange of static spin bosons between particles and holes,
which is a manifestation of the strong scattering due to preformed local moments in the bad metal regime.

While $\nabla^{\text{SBE},\ch}$ captures the asymptotic behavior of the charge vertex $f^\ch$, there are significant differences at small frequencies $\nu, \nu',\omega$.
The slices in the bottom left panel of Fig.~\ref{fig:U8} and panel (a) and (b) show that the attractive (blue) feature of $f^\ch$ has the opposite sign of $\nabla^{\text{SBE},\ch}$.
Apparently, in this case the SBE vertices cancel to some extent with the fully $U$-irreducible vertex $\varphi^{\firr,\ch}$.

Lastly, we note that the results obtained for $U/t=8$ and $\beta t=5$ correspond to a parameter regime
beyond the first divergence line of the $gg$-$ph$-irreducible two-particle self-energy~\cite{Schaefer13}.
In the parquet formalism this divergence cancels with the corresponding $gg$-$ph$-reducible vertex, leading to a \textit{finite} full vertex function $f$.
In contrast, in our results none of the SBE vertices $\nabla$ and hence neither the fully $U$-irreducible vertex $\varphi^{\firr}$
change their sign as a function of the interaction, which would indicate the crossing of a divergence line.

\section{Conclusions and outlook}~\label{sec:conclusions}
In this work we have expressed the four-point vertex function of the single-band Anderson impurity model in terms of three exchange bosons, which represent charge, spin, and singlet particle-particle fluctuations, respectively. 
On a formal level these bosons arise after classification of diagrams for the vertex into those that
are reducible with respect to the bare Hubbard interaction $U$ and those that are not.
We have shown that this classification leads to a parquet-like decomposition of the vertex into three $U$-\textit{reducible} and one fully $U$-\textit{irreducible} component.
However, this decomposition does not require the inversion of the Bethe-Salpeter equation and it is therefore not affected by divergences of the two-particle self-energy~\cite{Schaefer13,Chalupa18}.

Similar to the parquet decomposition, the classification of diagrams according to $U$-reducibility is
\textit{a priori} only a technical observation without physical implications.
However, it immediately reveals a direct physical interpretation. Since the Green's function lines can be contracted at a bare interaction vertex, the latter acts as a bosonic end-point where the fermions are coupled with collective excitations.
In a sense, the resulting philosophy is similar to the one behind the fluctuation exchange (FLEX) approximation~\cite{Bickers89},
where the vertex is however still computed from the two-particle self-energy, which is not a physical correlation function.
Our approach indeed realizes the concept of fluctuation exchange even more literally than FLEX, because the vertex is expressed in terms of observable effective bosons.

In fact, each $U$-reducible vertex represents the exchange of a \textit{single} boson during electronic scatterings.
We therefore refer to this decomposition as a single-boson exchange (SBE) decomposition of the full vertex and to the $U$-reducible components as SBE vertices.
The latter represent a crossing-symmetric subset of the vertex diagrams,
exact up to third order in the interaction [cf. Appendix~\ref{app:u2}], which gives rise to the vertex asymptotics.
Each SBE vertex is associated to characteristic scattering events,
which can be represented in terms of the Hedin three-leg vertex and the screened interaction.
As a matter of fact, only the fully $U$-irreducible vertex is intrinsically a four-point vertex which does not admit a representation in terms of single-boson exchange.

We have evaluated the components of the SBE decomposition for an Anderson impurity model, using the hybridization function which corresponds to the Hubbard model on the square lattice within the DMFT mapping.
As expected, the SBE vertices can be used to approximate the full vertex of the impurity in the weak-coupling regime
and for any interaction they provide a unified expression for the vertex asymptotics~\cite{Wentzell16,Kaufmann17}.
Remarkably, even for large interaction strength the dominant scattering processes are captured by the SBE vertices.
For example, in the bad metal regime they describe the strong suppression of charge propagation due to exchange of bosons of the spin channel between particles and holes.

The proposed SBE decomposition is a fairly general contribution to the diagrammatic theory of the vertex function, as such it is suggestive of a broad range of applications.
For example, in the future we will investigate the possibility to approximate the full vertex by the SBE vertices, neglecting the fully irreducible vertex, in order to calculate the DMFT
susceptibility~\cite{Georges96,Geffroy18,Krien19} and within diagrammatic extensions of DMFT~\cite{Rohringer17}.
A generalization to multi-orbital systems and simplified expressions for the SBE vertices may strongly ease the evaluation of vertex corrections to response functions and to the electronic self-energy, also in the context of electronic structure calculations~\cite{Held07}.
Furthermore, the utility of the SBE decomposition as an alternative to the parquet decomposition can be explored.
A related aspect that was not highlighted in this work is that the SBE equations can be formulated entirely in terms of three-leg vertices~\cite{Krien19-3}.
This makes the SBE approach crucially lighter from a computational point of view compared to the parquet equations that are formulated in terms of four-leg vertices.

\acknowledgments
We thank A. Toschi for his reading of the manuscript and many useful suggestions.
F.K. thanks D. Geffroy, J. Kaufmann, A.I. Lichtenstein and E.A. Stepanov for discussions.
M.C. acknowledges financial support from MIUR PRIN 2015 (Prot.2015C5SEJJ001) and SISSA/CNR project “Super- conductivity, Ferroelectricity and Magnetism in bad metals” (Prot. 232/2015).
A.V. acknowledges financial support from the Austrian Science Fund (FWF) through the Erwin Schr\"{o}dinger fellowship J3890-N36.
\appendix

\section{$\mathbf{U}$-$\mathbf{pp}$-irreducible vertices}\label{app:upp}
We separate $U$-$pp$-reducible diagrams from the vertex function.
We proceed in a similar way as in Ref.~\cite{Krien19}, where this is done for the particle-hole channel.

\subsection{Generalized susceptibility}
For the singlet channel one defines a generalized susceptibility as follows,
\begin{align}
    \chi^{\sing}_{\nu\nu'\omegap}=&-\beta g_\nu g_{\omegap-\nu}\delta_{\nu\nu'}+g_\nu g_{\omegap-\nu}\frac{1}{2}f^\sing_{\nu\nu'\omegap}g_{\nu'}g_{\omegap-\nu'},\label{app:gsuscsing}
\end{align}
where the singlet vertex function $f^\sing$ is given by equation~\eqref{eq:fsinglet},
cf. Ref.~\cite{Rohringer12}, but we choose a slightly different definition for $\chi^\sing$ because it leads to analogous relations as in the $ph$-channels.
$\chi^{\sing}$ satisfies the ladder equation,
\begin{align}
    \hat{\chi}^\sing_{\omegap}=&-\hat{\chi}^0_{\omegap}-\frac{1}{2}\hat{\chi}^0_{\omegap}\hat{\gamma}^\sing_{\omegap}\hat{\chi}^\sing_{\omegap},\label{app:sladder}
\end{align}
where $\gamma^{\sing}$ is the two-particle self-energy of the singlet-$pp$-channel ($gg$-$pp$-irreducible vertex) and
we adopted a matrix notation for $\chi^{\sing}_{\nu\nu'}(\omegap)$.
$\chi^0_{\nu\nu'}(\omegap)=\beta g_\nu g_{\omegap-\nu}\delta_{\nu\nu'}$ is the particle-particle bubble.
Note that matrix multiplication implies a factor $\beta^{-1}$.

As exercised in Ref.~\cite{Krien19} for the particle-hole channel, we like to define a $U$-$pp$-irreducible generalized susceptibility $\pi^{\sing}_{\nu\nu'\omegap}$
that does not include insertions of the bare interaction in the sense of the bottom diagram of Fig.~\ref{fig:ured}.
In the ladder equation~\eqref{app:sladder} all of the $U$-$pp$-reducible diagrams arise from the leading order of the two-particle self-energy $\gamma^{\sing}$,
which is $2U$ according to Ref.~\cite{Rohringer12}. We therefore define $\pi^{\sing}$ via the ladder equation,
\begin{align}
    \hat{\pi}^\sing_{\omegap}=&-\hat{\chi}^0_{\omegap}-\frac{1}{2}\hat{\chi}^0_{\omegap}\hat{\gamma}^{i,\sing}_{\omegap}\hat{\pi}^\sing_{\omegap},\label{app:piladder}
\end{align}
where $\gamma^{i,\sing}_{\nu\nu'}(\omegap)=\gamma^{\sing}_{\nu\nu'}(\omegap)-2U$. By inverting the matrix relations~\eqref{app:sladder} and~\eqref{app:piladder}
we arrive at the following relation between reducible and irreducible generalized susceptibility (see also Ref.~\cite{Krien19}),
\begin{align}
    \hat{\chi}^\sing_{\omegap}=&\hat{\pi}^{\sing}_{\omegap}+\frac{1}{2}\hat{\pi}^{\sing}_{\omegap}2U\hat{\chi}^\sing_{\omegap}.\label{app:xandpi1}
\end{align}
We write the matrix relation~\eqref{app:xandpi1} in explicit notation,
\begin{align}
    {\chi}^\sing_{\nu\nu'\omegap}=&{\pi}^{\sing}_{\nu\nu'\omegap}+\frac{1}{2}\left(\sum_{\nu_1}{\pi}^{\sing}_{\nu\nu_1\omegap}\right)
    U^{\sing}\left(\sum_{\nu_2}{\chi}^\sing_{\nu_2\nu'\omegap}\right),\label{app:xandpi}
\end{align}
where $U^\sing=2U$. This is the central relation between reducible and irreducible diagrams in the singlet-$pp$-channel.
We use it to derive several equations in the main text.

\subsection{Irreducible susceptibility and $\mathbf{pp}$-Hedin vertex}
Firstly, summing Eq.~\eqref{app:xandpi} over $\nu$ and $\nu'$ we arrive at a relation for the pairing susceptibility
$\chi^\sing_{\omegap}=-\left\langle \rho^-_{-\omegap}\rho^{+}_{\omegap}\right\rangle$ [$\rho^\pm$ are defined below Eq.~\eqref{eq:g3pp}],
\begin{align}
    \chi^{\sing}_{\omegap}=\sum_{\nu\nu'}{\chi}^\sing_{\nu\nu'\omegap}=&{\pi}^{\sing}_{\omegap}+\frac{1}{2}{\pi}^{\sing}_{\omegap}U^{\sing}\chi^{\sing}_{\omegap},\label{app:xfrompi}
\end{align}
where the polarization is defined as ${\pi}^{\sing}_{\omegap}=\sum_{\nu\nu'}{\pi}^{\sing}_{\nu\nu'\omegap}$.
We further define a $U$-$pp$-irreducible three-leg (or Hedin) vertex $\lambda^{\sing}$.
To this end, we sum equation~\eqref{app:xandpi} over $\nu'$,
\begin{align}
    \sum_{\nu'}{\chi}^\sing_{\nu\nu'\omegap}=&
    \sum_{\nu_1}{\pi}^{\sing}_{\nu\nu_1\omegap}\left(1+\frac{1}{2}U^{\sing}\sum_{\nu_2\nu'}{\chi}^\sing_{\nu_2\nu'\omegap}\right).
    \label{app:lambdadef1}
\end{align}
We identify $\sum_{\nu_2\nu'}{\chi}^\sing_{\nu_2\nu'\omegap}=\chi^{\sing}_{\omegap}$,
divide by $g_\nu g_{\omegap-\nu}(1+\frac{1}{2}U^{\sing}\chi^\sing_{\omegap})$, and define the right-sided Hedin vertex as,
\begin{align}
    \bar{\lambda}^{\sing}_{\nu\omegap}=&\frac{\sum_{\nu'}{\pi}^{\sing}_{\nu\nu'\omegap}}{g_\nu g_{\omegap-\nu}}
    =\frac{\sum_{\nu'}{\chi}^\sing_{\nu\nu'\omegap}}{g_\nu g_{\omegap-\nu}\left(1+\frac{1}{2}U^{\sing}{\chi}^\sing_{\omegap}\right)}.
    \label{app:lambdadef}
\end{align}
The generalized susceptibility is related to the three-point correlation function as $g^{(3),\sing}_{\nu\omegap}=\sum_{\nu'}{\chi}^\sing_{\nu\nu'\omegap}$,
leading to equation~\eqref{eq:lambdasing} in the main text. In a similar way one defines a left-sided Hedin vertex,
\begin{align}
    {\lambda}^{\sing}_{\nu\omegap}
    =&\frac{\sum_{\nu'}{\chi}^\sing_{\nu'\nu\omegap}}{g_\nu g_{\omegap-\nu}\left(1+\frac{1}{2}U^{\sing}\chi^{\sing}_{\omegap}\right)}.
    \label{app:lambdadefleft}
\end{align}

\subsection{$\mathbf{U}$-$\mathbf{pp}$-irreducible vertex function}\label{app:upp:firr}
We define analogous to equation~\eqref{app:gsuscsing} a vertex part $\varphi^{\pp,s}$ for the irreducible generalized susceptibility,
\begin{align}
    \pi^{\sing}_{\nu\nu'\omegap}=&\!-\!\beta g_\nu g_{\omegap-\nu}\delta_{\nu\nu'}\!+\!g_\nu g_{\omegap-\nu}\frac{1}{2}\varphi^{\pp,\sing}_{\nu\nu'\omegap}g_{\nu'}g_{\omegap-\nu'}.\label{app:gsuscsingirr}
\end{align}
We relate $\varphi^{\pp,\sing}$ to the full vertex $f^{\sing}$.
To this end, we insert Eqs.~\eqref{app:gsuscsing} and~\eqref{app:gsuscsingirr} into equation~\eqref{app:xandpi},
cancel $-\beta g_\nu g_{\omegap-\nu}\delta_{\nu\nu'}$ on both sides, and divide by $g_\nu g_{\omegap-\nu}g_{\nu'}g_{\omegap-\nu'}$,
\begin{align}
    f^\sing_{\nu\nu'\omegap}=&\varphi^{\pp,\sing}_{\nu\nu'\omegap}+
    \frac{\left(\sum_{\nu_1}{\pi}^{\sing}_{\nu\nu_1\omegap}\right)U^{\sing}\left(\sum_{\nu_2}{\chi}^\sing_{\nu_2\nu'\omegap}\right)}
    {g_\nu g_{\omegap-\nu}g_{\nu'}g_{\omegap-\nu'}},\label{app:fsing1}
\end{align}
a factor $\frac{1}{2}$ was canceled on both sides.
We combine this relation with the definitions of the Hedin vertices in Eqs.~\eqref{app:lambdadef} and~\eqref{app:lambdadefleft},
\begin{align}
    f^\sing_{\nu\nu'\omegap}=&\varphi^{\pp,\sing}_{\nu\nu'\omegap}
    +\bar{\lambda}^{\sing}_{\nu\omegap}U^{\sing}\left(1+\frac{1}{2}U^{\sing}\chi^{\sing}_{\omegap}\right){\lambda}^{\sing}_{\nu'\omegap}\label{app:firrsing}.
\end{align}
Finally, we identify the screened interaction,
\begin{align}
    w^{\sing}_{\omegap}=U^{\sing}\left(1+\frac{1}{2}U^{\sing}\chi^{\sing}_{\omegap}\right),
\end{align}
which extends equation~\eqref{eq:w} to the case $\alpha=\sing$.
Furthermore, Eq.~\eqref{app:firrsing} is the desired decomposition~\eqref{eq:us} of the vertex into irreducible and reducible parts.

\subsection{$\mathbf{U}$-$\mathbf{pp}$-reducible diagrams in particle-hole notation}\label{app:upp:nablapp}
In the main text we focus on the SBE decomposition of the full vertex function $f^{\ch/\sz}$ in particle-hole notation,
however, equation~\eqref{app:firrsing} is formulated in the particle-particle notation.
In order to obtain the $U$-$pp$-reducible part of $f^{\ch/\sz}$ we use the relations~\eqref{eq:fsinglet} and~\eqref{eq:triplet}
between particle-hole and particle-particle notations, leading to,
\begin{align}
    f^{\ch/\sz}_{\nu\nu'\omega}=&f^{\trip}_{\nu\nu',\omega+\nu+\nu'}\pm\frac{1}{2}(f^{\sing}_{\nu\nu',\omega+\nu+\nu'}+f^{\trip}_{\nu\nu',\omega+\nu+\nu'}),\notag
\end{align}
where $\sing$ and $\trip$ denote the singlet and triplet channel.
We insert the decomposition~\eqref{eq:us} of the singlet vertex $f^{\sing}$ and obtain the desired relation for the particle-hole notation,
\begin{align}
    f^{\ch/\sz}_{\nu\nu'\omega}=&\varphi^{\pp,\ch/\sz}_{\nu\nu',\omega+\nu+\nu'}
    \pm\frac{1}{2}\bar{\lambda}^{\sing}_{\nu,\omega+\nu+\nu'}w^\sing_{\omega+\nu+\nu'}\lambda^{\sing}_{\nu',\omega+\nu+\nu'},\notag
\end{align}
which is equation~\eqref{eq:uspp} in the main text and we summarized all $U$-$pp$-irreducible contributions into the vertex,
\begin{align}
    \varphi^{\pp,\ch/\sz}_{\nu\nu'\omegap}=&f^{\trip}_{\nu\nu'\omegap}\pm\frac{1}{2}(\varphi^{pp,\sing}_{\nu\nu'\omegap}+f^{\trip}_{\nu\nu'\omegap}).
    \label{app:phising}
\end{align}
Note that $\varphi^{pp,\sing}$ is the $U$-$pp$-irreducible vertex of the singlet channel,
whereas the triplet vertex $f^{\trip}$ is $U$-$pp$-irreducible by construction [see comments below equation~\eqref{eq:triplet}]
and must hence be fully accounted to the $U$-$pp$-irreducible parts of $f^\ch$ and $f^{\sz}$, respectively.

\section{Equivalence of left- and right-sided Hedin vertex}\label{app:upp:lambdasym}
We relate the left- and right-sided Hedin vertices $\lambda^{\alpha}$ and $\bar{\lambda}^{\alpha}$.
To do this, we use the following relation for the vertex function that holds under time-reversal symmetry combined with SU($2$) symmetry~\cite{Rohringer12},
\begin{align}
    f^\alpha_{\nu\nu'\omega}=&f^\alpha_{\nu'\nu\omega},\;\;\alpha=\ch,\sz,\sing.\label{app:fsingsym}
\end{align}
We combine this with equations~\eqref{eq:fph} and~\eqref{eq:uspp} to obtain a relation for the irreducible vertices $\varphi$,
\begin{align}
    \varphi_{\nu\nu'\omega}+\bar{\lambda}_{\nu\omega}w_\omega\lambda_{\nu'\omega}
    =&\varphi_{\nu'\nu\omega}+\bar{\lambda}_{\nu'\omega}w_\omega\lambda_{\nu\omega},\label{app:phisymph}
\end{align}
where we temporarily dropped the labels $ph, pp,$ and $\alpha$.
For the particle-particle channel ($\alpha=\sing$) the bosonic frequency in Eqs.~\eqref{app:fsingsym} and~\eqref{app:phisymph} is that of a particle-particle pair, $\omega\rightarrow\omegap$.
\begin{figure}
\begin{center}
      \includegraphics[width=.48\textwidth]{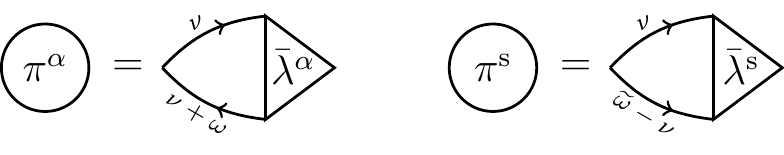}
\end{center}
    \caption{\label{fig:polarization}
    Relation between the polarization and the right-sided Hedin vertex $\bar{\lambda}$
    (similar relations hold for the left-sided $\lambda$ by attaching legs from the right).
    (Left) Particle-hole channels. (Right) Singlet-particle-particle channel.
    }
    \end{figure}

\subsection{Particle-hole channel}\label{app:sym:ph}
According to Ref.~\cite{Krien19} the $U$-$ph$-irreducible vertex $\varphi^{ph,\alpha}$ is related to the right- and left-sided Hedin vertex as~\footnote{
    In Ref.~\cite{Krien19} the $U$-$ph$-irreducible vertex $\varphi^{ph}$ is called $f^{i}$.},
\begin{align}
    \bar{\lambda}^{\alpha}_{\nu\omega}=&1+\sum_{\nu'}\varphi^{\ph,\alpha}_{\nu\nu'\omega}g_{\nu'}g_{\nu'+\omega},\label{app:lambdafromphiright}\\
        {\lambda}^{\alpha}_{\nu\omega}=&1+\sum_{\nu'}g_{\nu'}g_{\nu'+\omega}\varphi^{\ph,\alpha}_{\nu'\nu\omega}.\label{app:lambdafromphileft}
\end{align}
We make use of these relations in equation~\eqref{app:phisymph} by multiplying the latter by $g_{\nu'}g_{\nu'+\omega}$, summing over $\nu'$,
and adding $1$ on both sides, leading to,
\begin{align}
    &\bar{\lambda}^\alpha_{\nu\omega}+\bar{\lambda}^\alpha_{\nu\omega}w^\alpha_\omega\left(\sum_{\nu'}\lambda^\alpha_{\nu'\omega}g_{\nu'}g_{\nu'+\omega}\right)\notag\\
    =&\lambda^\alpha_{\nu\omega}+\left(\sum_{\nu'}g_{\nu'}g_{\nu'+\omega}\bar{\lambda}^\alpha_{\nu'\omega}\right)w^\alpha_\omega\lambda^\alpha_{\nu\omega}.\label{app:phisymphsummed}
\end{align}
The sums yield the polarization~\cite{Krien19},
$\sum_{\nu'}\lambda^\alpha_{\nu'\omega}g_{\nu'}g_{\nu'+\omega}=\sum_{\nu'}g_{\nu'}g_{\nu'+\omega}\bar{\lambda}^\alpha_{\nu'\omega}=\pi^\alpha_\omega$,
see also left-hand-side of Fig.~\ref{fig:polarization},
\begin{align}
    \bar{\lambda}^\alpha_{\nu\omega}(1+w^\alpha_\omega\pi^\alpha_\omega)=&\lambda^\alpha_{\nu\omega}(1+\pi^\alpha_\omega w^\alpha_\omega),
\end{align}
and hence,
\begin{align}
    \bar{\lambda}^\alpha_{\nu\omega}=&\lambda^\alpha_{\nu\omega},\label{app:leftrighteq}
\end{align}
for $\alpha=\ch,\sz$ and right- and left-sided Hedin vertex of the particle-hole channel are equivalent.

\subsection{Particle-particle channel}\label{app:sym:pp}
For the particle-particle case we obtain the corresponding relation between Hedin vertex and $U$-$pp$-irreducible vertex $\varphi^{pp,\sing}$
from Eqs.~\eqref{app:gsuscsingirr} and~\eqref{app:lambdadef},
\begin{align}
    \bar{\lambda}^\sing_{\nu\omegap}=&-1+\frac{1}{2}\sum_{\nu'}\varphi^{\pp,\sing}_{\nu\nu'\omegap}g_{\nu'}g_{\omegap-\nu'}\label{app:lambdafromphirightpp}\\
        {\lambda}^\sing_{\nu\omegap}=&-1+\frac{1}{2}\sum_{\nu'}g_{\nu'}g_{\omegap-\nu'}\varphi^{\pp,\sing}_{\nu'\nu\omegap}.\label{app:lambdafromphileftpp}
\end{align}
Starting from Eq.~\eqref{app:phisymph} for $\varphi^{pp,\sing}$,
similar steps as in the particle-hole case show the equivalence,
\begin{align}
    \bar{\lambda}^\sing_{\nu\omegap}={\lambda}^\sing_{\nu\omegap}.\label{app:leftrighteqpp}
\end{align}
Notice that in this case the polarization $\pi^\sing_{\omegap}$ is given as on the right-hand-side of Fig.~\ref{fig:polarization}, see also Eq.~\eqref{app:lambdadef}.

\subsection{Complex conjugation}
In Secs.~\ref{app:sym:ph} and~\ref{app:sym:pp} we made use of the time-reversal symmetry,
however, the right- and left-sided Hedin vertex are also connected by a relation of the vertex function to its complex conjugate~\cite{Rohringer12},
\begin{align}
    f^\alpha_{\nu\nu'\omega}=&\left(f^\alpha_{-\nu',-\nu,-\omega}\right)^*,
    \label{app:fsingconj}
\end{align}
which is a weaker assumption than~\eqref{app:fsingsym}. Using equation~\eqref{app:fsingconj} it is easy to prove,
\begin{align}
\bar{\lambda}^\alpha_{\nu\omega}=(\lambda^\alpha_{-\nu,-\omega})^*.
\end{align}
When also Eqs.~\eqref{app:leftrighteq} and~\eqref{app:leftrighteqpp} are valid this implies the symmetry relation,
$\bar{\lambda}^\alpha_{\nu\omega}=(\bar{\lambda}^\alpha_{-\nu,-\omega})^*$.
\begin{figure}
    \begin{center}
      \includegraphics[width=.48\textwidth]{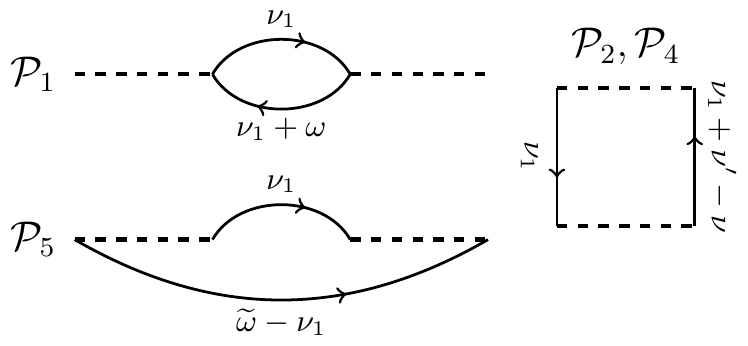}
\end{center}
    \caption{\label{fig:u2diagrams}
    $U$-reducible diagrams at order $\mathcal{O}(U^2)$. 
    The vertex $\nabla^{ph}$ generates the diagram $\mathcal{P}_1$,
    $\nabla^{\overline{ph}}$ generates $\mathcal{P}_2$ and $\mathcal{P}_4$,
    and $\nabla^{pp}$ yields diagram $\mathcal{P}_5$, see text.
    Labels $\mathcal{P}$ as in Ref.~\cite{Rohringer12}.
    }
    \end{figure}

\section{Lowest-order $U$-reducible diagrams}\label{app:u2}
We demonstrate that already at order $\mathcal{O}(U^2)$ the $U$-reducible vertices $\nabla^{ph}$, $\nabla^{\overline{ph}}$,
and $\nabla^{pp}$ of the SBE-decomposition~\eqref{eq:jib_full} generate topologically distinct diagrams.
In the following we consider order $\mathcal{O}(U)$, $\mathcal{O}(U^2)$, and $\mathcal{O}(U^3)$.
We denote diagrams as in Ref.~\cite{Rohringer12} (although we use a dashed line to depict the bare interaction):

\textit{First order.} At order $\mathcal{O}(U)$ there is only one diagram, the bare interaction.
This diagram arises as the leading order of all three $U$-reducible vertices
$\nabla^{{ph},\alpha}, \nabla^{\overline{ph},\alpha},$ and $\nabla^{\overline{pp},\alpha}$.
To see this, we consider the leading order of the screened interaction~\eqref{eq:w}
and of the Hedin vertices~\eqref{app:lambdafromphiright} and~\eqref{app:lambdafromphirightpp},
\begin{align}
    w^\alpha&=U^\alpha+... ,\notag\\
    \bar{\lambda}^{\ch/\sz}&=\phantom{ + {}} 1+... ,\\
    \bar{\lambda}^{\sing}&=-1+...
\end{align}
This yields for the $U$-$ph$-reducible vertex~\eqref{eq:nablahph},
\begin{align}
    \nabla^{ph,\alpha}=&U^\alpha+\mathcal{O}(U^2).
\end{align}
We insert this result into the $U$-$\overline{ph}$-reducible vertex~\eqref{eq:nablavph},
\begin{align}
    \nabla^{\overline{ph},\alpha}
    =&-\!\frac{1}{2}\!\left(U^\ch+[3-4\delta_{\alpha,\sz}]U^\sz\right)+\mathcal{O}(U^2)\notag\\
    =&U^\alpha+\mathcal{O}(U^2),
\end{align}
where we used $U^\ch=+U$ and $U^\sz=-U$. Finally, we consider the $U$-$pp$-reducible vertex~\eqref{eq:nablapp},
\begin{align}
    \nabla^{pp,\alpha}=&\frac{1-2\delta_{\alpha,\sz}}{2}2U+\mathcal{O}(U^2)=U^\alpha+\mathcal{O}(U^2),
\end{align}
where $\alpha=\ch,\sz$ and we inserted the bare interaction of the singlet channel~\eqref{eq:ubarepp}, $U^{\sing}=2U$.
As expected, the $U$-reducible vertices all yield the bare interaction $U^\alpha$ as the leading order.
In the SBE-decomposition~\eqref{eq:jib_full} it is therefore necessary to subtract it two times in order to avoid double-counting.
\begin{figure}
    \begin{center}
      \includegraphics[width=.48\textwidth]{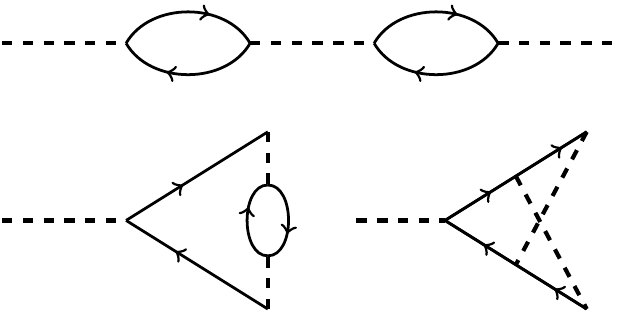}
\end{center}
    \caption{\label{fig:o3}
    $U$-$ph$-reducible diagrams of order $\mathcal{O}(U^3)$.
    The triangular shapes in the bottom diagrams are parts of the left-sided Hedin vertex $\lambda^{\ch/\sz}$.
    At order $\mathcal{O}(U^3)$ all diagrams are $U$-reducible in some way.
    (Bottom left) $U$-$\overline{ph}$-reducible vertex correction. (Bottom right) $U$-$pp$-reducible vertex correction.
    }
    \end{figure}
    
\textit{Second order.} We consider the diagrams of order $\mathcal{O}(U^2)$ beginning with the particle-hole channels.
At this order all diagrams are $U$-reducible and arise once again from $\bar{\lambda}^{\ch,\sz}\approx1$,
whereas the screened interaction~\eqref{eq:w} is expanded to order $\mathcal{O}(U^2)$.
For this it is useful to write the screened interaction for the particle-hole channel as~\cite{Krien19},
\begin{align}
    w^\alpha_\omega=&\frac{U^\alpha}{1-U^\alpha\pi^\alpha_\omega}=U^\alpha+U^\alpha\pi^\alpha_\omega U^\alpha+...\notag\\
    =&U^\alpha+U^{2}\sum_{\nu_1}g_{\nu_1}g_{\nu_1+\omega}+\mathcal{O}(U^3).\label{app:wexpand}
\end{align}
In the first line we expanded the geometric series, in the second line we used the relation between polarization $\pi$ and Hedin vertex $\bar{\lambda}$
depicted on the left-hand-side of Fig.~\ref{fig:polarization} and truncated $\bar{\lambda}$ after the leading term $1$.
We arrive at the following diagram for the $U$-$ph$-reducible vertex
in Eq.~\eqref{eq:nablahph},
\begin{align}
    \mathcal{P}_1=U^{2}\sum_{\nu_1}g_{\nu_1}g_{\nu_1+\omega},
\end{align}
and there is no difference between the charge and spin channels.
$\mathcal{P}_1$ is indeed the same diagram as in Ref.~\cite{Rohringer12}, the only $gg$-$ph$-reducible diagram at order $\mathcal{O}(U^2)$.

We further use $\mathcal{P}_1$ to evaluate the $U$-$\overline{ph}$-reducible vertex $\nabla^{\overline{ph},\alpha}$ in equation~\eqref{eq:nablavph},
\begin{align}
    \mathcal{P}_2+(1-2\delta_{\alpha,\sz})\mathcal{P}_4=-2(1-\delta_{\alpha,\sz})U^{2}\sum_{\nu_1}g_{\nu_1}g_{\nu_1+\nu'-\nu}.
\end{align}
$\mathcal{P}_2$ and $\mathcal{P}_4$ are $gg$-$\overline{ph}$-reducible (the transferred momentum of the particle-hole bubble is $\nu'-\nu$)
and they cancel each other in the spin channel, as in Ref.~\cite{Rohringer12}. Also the negative prefactor is recovered correctly.

Finally, we come to the particle-particle channel, where we truncate the Hedin vertex after the leading order, $\bar{\lambda}^{\sing}\approx-1$ .
Due to choice of definitions the screened interaction $w^\sing$ is related to the polarization $\pi^\sing$
in a slightly different way than in the particle-hole channel, cf. Eq.~\eqref{app:wexpand},
\begin{align}
    w^{\sing}_{\omegap}=&\frac{U^{\sing}}{1-\frac{1}{2}U^{\sing}\pi^\sing_{\omegap}}=U^{\sing}+U^{\sing}\frac{1}{2}\pi^{\sing}_{\omegap}U^{\sing}+...\label{app:wexpandpp}\\
    =&2U-\frac{1}{2}(2U)^{2}\sum_{\nu_1}g_{\nu_1}g_{\omegap-\nu_1}+\mathcal{O}(U^3).\notag
\end{align}
In the second step we inserted the bare interaction $U^\sing=2U$
and truncated the Hedin vertex $\bar{\lambda}^\sing\approx-1$ in the relation for the polarization $\pi^{\sing}$ on the right-hand-side of Fig.~\ref{fig:polarization}.
We obtain the following diagram for the $U$-$pp$-reducible vertex in Eq.~\eqref{eq:nablapp},
\begin{align}
    \mathcal{P}_5
    =&-(1-2\delta_{\alpha,\sz})U^2\sum_{\nu_1}g_{\nu_1}g_{\omegap-\nu_1}.
\end{align}
The prefactor of this diagram is consistent with Ref.~\cite{Rohringer12}, note that $\omegap=\omega+\nu+\nu'$.

The diagrams $\mathcal{P}_1, \mathcal{P}_2, \mathcal{P}_4$, and $\mathcal{P}_5$ are shown in Fig.~\ref{fig:u2diagrams}.
The $U$-reducible vertices $\nabla$ indeed generate only one identical diagram, the bare interaction,
whereas from order $\mathcal{O}(U^2)$ onwards one is left with topologically distinct diagrams: $\nabla^{ph},\nabla^{\overline{ph}}$, and $\nabla^{pp}$ generate 
$gg$-$ph$-, $gg$-$\overline{ph}$, and $gg$-$pp$-reducible diagrams, respectively,
there is hence no further double-counting.

\textit{Third order.} We briefly discuss $\mathcal{O}(U^3)$-diagrams.
According to Ref.~\cite{Rohringer12} the first fully $gg$-irreducible diagram beyond the bare interaction
is the envelope diagram of order $\mathcal{O}(U^4)$ on the left of Fig.~\ref{fig:firr}.
Therefore, at order $\mathcal{O}(U^3)$ only the bare interaction can be a fully $gg$-irreducible building block.
However, three bare interaction lines can not be connected to form a fully $U$-irreducible diagram, therefore, also at this order all diagrams are $U$-reducible.
We illustrate this in Fig.~\ref{fig:o3}, which shows three diagrams of order $\mathcal{O}(U^3)$ which contribute to the $U$-$ph$-reducible vertex $\nabla^{ph}$.
The first fully $U$-irreducible diagrams arise at expansion order $\mathcal{O}(U^4)$, see also Fig.~\ref{fig:firr}.

 \bibliography{main}

\end{document}